\documentclass[11pt,preprint]{aastex}

\newcommand{\lam}{$\lambda$}

\renewcommand{\ion}[2]{#1\,{\sc #2}}
\newcommand{\kms}{km~s$^{-1}$}

\newcommand{\lya}{Ly$\alpha$}

\begin{document}


\title{Forbidden and intercombination lines of RR Telescopii:
  wavelength measurements and energy levels}


\author{P. R. Young\altaffilmark{1}, 
 U. Feldman\altaffilmark{2} \and A. Lobel\altaffilmark{3}}

\altaffiltext{1}{George Mason University, 4400 University Drive, Fairfax, VA 22030}
\altaffiltext{2}{Artep Inc., Ellicott City, MD 21042}
\altaffiltext{3}{Royal Observatory of Belgium, Ringlaan 3, B-1180
  Brussels, Belgium}


\begin{abstract}
Ultraviolet and visible spectra of the symbiotic nova RR Telescopii
are used to derive reference wavelengths for many forbidden
and intercombination transitions of ions +1 to +6 of elements C, N, O,
Ne, Na, Mg, Al, Si, P, S, Cl, Ar, K and Ca. The wavelengths are then
used to determine new energy values for the levels within the ions'
ground configurations or first excited configuration. The spectra were measured by the
Space Telescope Imaging Spectrograph of the Hubble Space Telescope and the UltraViolet
Echelle Spectrograph of the European Southern Observatory in 2000 and
1999, respectively, and cover 1140 to 6915~\AA. Particular care was taken to assess the accuracy
of the wavelength scale between the two instruments. An investigation
of the profiles of the emission lines reveals that the nebula consists
of at least two
plasma components at different velocities. The components have
different densities, and a simple model of the lines' emissions
demonstrates that most of the lines principally arise from the high
density component. Only these lines were used for the wavelength study.
\end{abstract}


\section{Introduction}

Forbidden lines of ions from low ionization stages having very long
decay rates can only be measured in astronomical plasma 
sources as it is not possible to attain the necessary, very low plasma
densities in laboratory plasma sources. For 'coronal' ions, i.e.,
those with typically six or more electrons removed, it is possible to
measure the forbidden lines in solar spectra taken above the solar
limb and measurements obtained from Skylab and SOHO spectra are
presented in \citet{doschek76}, \citet{doschek77}, \citet{sandlin77} and \citet{feldman07}. For lower
ionization stages it is necessary to observe other astronomical
sources and, in particular, nebulae. The wavelengths derived by
\citet{bowen60} are for many forbidden lines still the standard
references used in astronomy and the present work seeks to update
these values using emission line measurements of the nova RR
Telescopii obtained with space-based and ground-based
spectrographs. Of particular interest are forbidden lines observed
below 3000~\AA\ that were inaccessible at the time of Bowen's work.

RR Tel has long been a favorite target of spectroscopists on account
of its very rich emission line spectrum, and indeed updates to some of
the Bowen wavelengths were provided by \citet{thackeray77} and
\citet{penston83} using visible and ultraviolet spectra of RR Tel. The
emission lines arise from a nebula that envelopes a late-type giant
and hot white dwarf, and the system is classified as a symbiotic nova
that went into outburst in 1944. The white dwarf
temperature was determined to be around 142,000~K from X-ray
observations \citep{jordan94} which is sufficient to produce
ionization 
stages up to +6 in the nebula. Of great value for ultraviolet
observations of RR Tel is the low extinction along the line-of-sight
towards the system, with $E(B-V)$ values of between 0 and 0.10
determined by different methods \citep{young05a,selvelli07}, which
ensures strong signals in short wavelength lines. In 
addition, the interstellar absorption lines are weak in the spectrum
and redshifted relative to the system's radial velocity of $-62$~\kms\
\citep{thackeray77}.

In the present work new wavelengths for 88 forbidden and
intercombination lines belonging
to ion species with charge states between +1 and +6 are
presented. Careful attention is paid to deriving an accurate rest
wavelength scale for both the UV and visible spectra by using lines
from low ionization stages, and to deriving accurate error
estimates for the wavelengths. The wavelengths are then used in
Sect.~\ref{sect.energy} to derive new energy level values for the ions.

\section{Observations}

The observations analyzed here were obtained with the Space Telescope
Imaging Spectrograph (STIS) on board the Hubble Space Telescope (HST),
and with the UVES \citep{dekker00} echelle spectrometer installed at the
Kueyen telescope of the Very Large Telescope (VLT). These data have
been used previously by  \citet{selvelli00}, \citet{keenan02},
\citet{young05a}, \citet{skopal07} 
and \citet{selvelli07} where more details can be found. We summarize
the main details here.

The STIS observations were obtained on 2000 October 18 and yielded
complete spectra over the wavelength range 1140--7051~\AA. The visible
spectra in the range 3022--7051~\AA\ were obtained with the STIS CCD
and are of low resolution so are not suitable for accurate wavelength
measurements. The UV spectra in the range 1140--3120~\AA\ have a high
resolution of 30,000--45,000, and were obtained with three exposures. The
first used the E140 echelle grating to cover the wavelength range 1140
to 1730~\AA, while the following two exposures used the E230M grating
to observe the wavelength ranges 1606--2367 and 2274--3120~\AA. We
will refer to the three STIS exposures as the short, medium and long
wavelength (SW, MW and LW) exposures. The
data were downloaded from the MAST archive which delivers calibrated
1D spectra. The data analyzed here had been processed with version~2.22
of CALSTIS, the STIS calibration pipeline.

The UVES spectra were obtained in 1999 October and cover two
wavelength ranges: 3085--3914~\AA\ and 4730--6915~\AA. The spectra
were reduced using the MIDAS package by one of the authors (A.~Lobel)
and have been corrected for the Earth's motion using a velocity of
$-24.6$~\kms\ \citep{selvelli00}.

\section{Plasma components}\label{sect.cpts}

\begin{figure}[t]
\epsscale{0.8}
\plotone{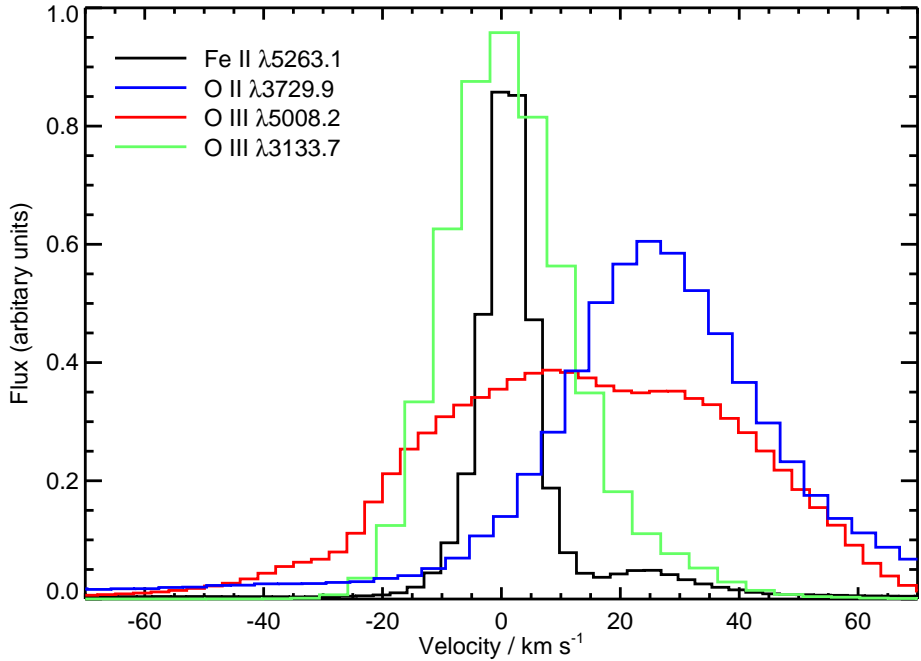}
\caption{Comparison of line profiles found in the UVES spectrum. The
  velocity scale has been set to the radial velocity of the system, as
determined from \ion{Fe}{ii} lines (see text).}
\label{fig.uves.cpts}
\end{figure}

The aim of the present work is to derive updated reference wavelengths
for forbidden and intercombination transitions of ionized atoms from
the RR Tel spectra. This is achieved by first deriving a reference
wavelength scale using emission lines with accurately-known
wavelengths. A key assumption in the work is that all of the emission
lines are emitted from a plasma with the same line-of-sight velocity
such that the results are not affected by relative Doppler shifts of
different species.

Inspection of certain line profiles (Fig.~\ref{fig.uves.cpts}),
however, clearly demonstrates that the RR Tel nebula has at least two distinct
plasma components with different velocities. This was first noticed by
\citet{schild97} who obtained high resolution line profiles for the
density sensitive \ion{O}{iii} \lam5007 and \lam4363 forbidden lines.  \citet{schild97} suggested that
the \ion{O}{iii} emission can be assigned to a high density component
at the radial velocity of the star, and a low density component
blueshifted relative to this component by 20~\kms. 
The \ion{O}{iii} \lam4363 line profile presented by \citet{crawford99}
also demonstrated `rest' and blueshifted components, although the
blueshift was 28~\kms\ in this case. The authors also presented line
profiles from four other emission lines that suggested up to four distinct
velocity components in the system. Note that some caution should be
attached to the results of these two papers as the \citet{schild97}
work was a brief conference proceedings report with no discussion on how
the velocity scale was derived, while \citet{crawford99} also did not
give details about the derivation of the velocity scale.

The spectra from UVES presented here reveal a different velocity
structure than those presented by \citet{schild97} and
\citet{crawford99}, and which is best illustrated by the line profiles
presented  in Fig.~\ref{fig.uves.cpts}. Four emission line profiles
are shown, and the velocity scale has been corrected for the average
velocity of the \ion{Fe}{ii} forbidden lines
(Sect.~\ref{sect.uves}). The \ion{Fe}{ii} velocity corresponds to the
classical radial velocity of the RR Tel nebula
\citep{thackeray50,thackeray77}, and we term this the `rest' component
of the nebula. The four profiles belong to  three forbidden lines,
\ion{Fe}{ii} \lam5263.1, \ion{O}{ii} 3729.9 and \ion{O}{iii}
\lam5008.2, and the resonance line \ion{O}{iii} 3133.7 which is
actually fluoresced through the \ion{He}{ii} \lam304 EUV line and so
is not optically thick.
\ion{O}{ii} \lam3729 reveals a broad, nearly
symmetric Gaussian that is redshifted by $\approx 20$~\kms. A
comparison of the strength of this line with the nearby \ion{O}{ii} \lam3727.1
(which lies partly in the wing of \ion{Ca}{vi} 3726.5, but can be
accurately estimated) suggests a density of $< 10^3$~cm$^{-3}$ using
the atomic model from the CHIANTI atomic database
\citep{dere97,dere09}. This is consistent with \lam3729.9 being formed
in the low density plasma component found by \citet{schild97}, however
the velocity shift is in the opposite direction. We speculate that
there may be some time dependence to the velocity of the low density
plasma component. Note that \citet{crawford99} presented a line
profile of \ion{O}{ii} \lam4072.2 which showed a two component
structure. This is a recombination line and so is formed in the
O$^{2+}$ emitting region.

\ion{O}{iii} \lam5008.2 shows a complex structure  suggesting three
plasma components: one at around
$+20$~\kms\ consistent with the \ion{O}{ii} line, another at around
$+10$~\kms, and a further one at around $-30$ to $-40$~\kms. It is
somewhat similar to the profiles presented by \citet{schild97} and
\citet{crawford99} however the main body of the profile here lies near
the rest velocity of the system whereas in these two works it was
blueshifted by 20--30~\kms. \ion{O}{iii} \lam3133 shows a strong
component at the rest velocity of the system, with an extended wing on
the long wavelength side of the profile, which may indicate a small
contribution from the low density plasma component. A similar profile is
found for a number of strong intercombination lines in the STIS spectrum, including
\ion{C}{iii} \lam1908.7, \ion{O}{iii} \lam1666.2, \ion{O}{iv}
\lam1401.2 and \ion{N}{iv} 1486.5.

Finally we show \ion{Fe}{ii} \lam5263.1 which is one of the forbidden
lines used to determine the RR Tel radial velocity. The profile
actually consists of two components, the stronger is the one used to
determine the radial velocity, while the weaker is close in velocity
to the \ion{O}{ii} \lam3729 line. 

The different velocity structure shown by these line profiles presents
a fundamental problem: how can an RR Tel emission line be used to
determine a rest wavelength for an atomic transition if there is
uncertainty over which plasma component the line arises from?
Appendix~\ref{app.model} presents a
simple model of the RR Tel nebula represented by  low density and
high density plasma components. Using the emission models from CHIANTI it is
demonstrated that only a small number of emission lines are expected
to have 
significant emission from the low density, red-shifted plasma
component (which gives rise to \ion{O}{ii} \lam3729.9). The vast majority
of the lines studied in the present work predominantly arise from the
rest component of the plasma (which gives rise to the strong, narrow
component of the \ion{Fe}{ii} \lam5263.1 line). 

\begin{figure}[h]
\plotone{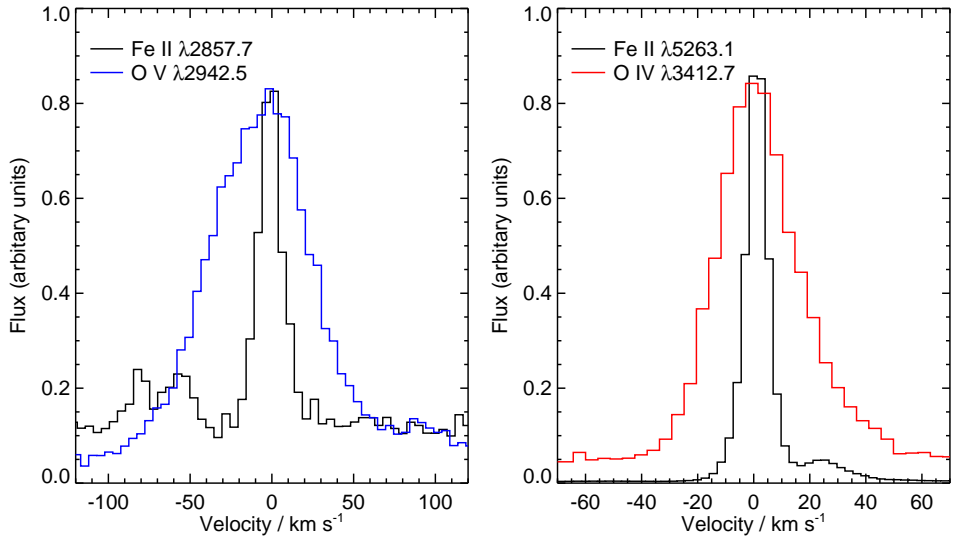}
\caption{A comparison of high ionization line profiles with profiles
  for \ion{Fe}{ii} wavelength fiducial lines. The left panel shows two
profiles from the STIS spectra, the right panel two profiles from the
UVES spectra. The velocity scale has been set to the radial velocity
of the system as determined from the wavelength fiducial lines (see Sect.~\ref{sect.abs.wvl}).}
\label{fig.oxy.rec}
\end{figure}

The emission lines shown in Fig.~\ref{fig.uves.cpts} are from low
ionization species, yet many of the lines studied in the present work
are from higher ionization species. Therefore another uncertainty is
whether further plasma components become apparent for higher
ionizations that are not revealed in the lower ionization lines.
This can be studied
by considering two lines of \ion{O}{iv} and \ion{O}{v}, formed
through recombination onto O$^{4+}$ (IP=77.4~eV) and O$^{5+}$
(IP=113.9~eV). Fig.~\ref{fig.oxy.rec} compares the line profiles of
\ion{O}{iv} \lam3412.67 and \ion{O}{v} \lam2942.51 with two of the 
\ion{Fe}{ii} emission lines used to determine the STIS and UVES rest
wavelength scales. The \ion{O}{iv} line is the $3p$ $^2P_{3/2}$ --
$3d$ $^2D_{5/2}$ transition, while \ion{O}{v} is $5g$ $^{1,3}G$ --
$6h$ $^{1,3}H$ (an unresolved multiplet). Laboratory wavelengths for
these transitions were measured by \citet{bromander69} and
\citet{bockasten68}, respectively, to accuracies of 0.05~\AA\
(5~\kms). The line profiles shown in Fig.~\ref{fig.oxy.rec} show that
the line centroids of the two high ionization oxygen lines are in
excellent agreement with the \ion{Fe}{ii} lines, confirming that the
high ionization species are emitted from the rest component of the RR
Tel nebula plasma.

\section{Time variability}

As the two sets of spectra considered in the present work are
separated by a year in time, we briefly discuss the time variability
of the RR Tel emission line spectra. RR Tel has had only one known
outburst, in 1944, and has evolved rather slowly since then.
\citet{thackeray77} performed a
detailed analysis of the variability of the 
visible spectrum  over the period 1951--1973, and Fig.~8 of this work
illustrates the increasing levels of ionization in the nebula for iron
over
this time. From 1954 to around 1960, the sequence of iron ions from
\ion{Fe}{iii} to \ion{Fe}{vii} progressively peaked and faded, with
\ion{Fe}{vii} (ionization potential, IP, of 99.1~eV) remaining strong
at the end of the observation period. The ion with the highest IP,
\ion{Ca}{vii} with a value of 108.8~eV, only became apparent in
1968. Although the low ionization species generally declined in
strength, they remained present in the spectrum.

RR Tel was regularly observed by IUE, and \citet{zuccolo97} presented
measurements of emission line fluxes from 1978 to 1993. The level of
ionization continued to increase since the work of
\citet{thackeray77}, as best indicated by the emergence of UV lines of
\ion{Mg}{vi} (IP=141.3~eV) in 1983 which increased in strength by a
factor three up to 1993. Comparing with the STIS spectrum of 2000, we
find that all of the  emission lines in the 1140--1900~\AA\ range have
become weaker since 1993, with the lower ionization stages (e.g.,
\ion{Fe}{ii}, \ion{O}{i}) by around a factor 10, moderate ionization
stages (e.g., \ion{O}{iv}, \ion{Si}{iv}) by a factor of a few, and the
highest ionization stage (\ion{Mg}{vi}) by only 40\%.

\citet{kotnik06} have demonstrated that RR Tel has undergone three
dust obscuration events, with the most recent in 1996--2000 which
coincided with a decline in the visual brightness of RR
Tel. \citet{kotnik09} presented measurements of optical emission lines
from 1996 and 2000 and found that all emission lines have declined in
flux, with the lower ionization species showing larger falls and the
higher ionization species showing smaller falls. This appears to be
consistent with the change in ultraviolet emission line fluxes between
1993 and 2000.

No evidence has been presented in the literature for significant short
term (timescales of 1 year) changes in the spectrum of RR Tel since
the 1944 outburst and, additionally, \citet{zuccolo97} found no
significant changes in emission line centroids with time for the IUE
observations of 1978--1993. It is thus reasonable to assume that the
STIS and UVES spectra can be combined to determine rest wavelengths
for the system, however the radial velocity discrepancy discussed in 
Sect.~\ref{sect.stis} may point to a small anomaly between the
two observations.

\section{Absolute wavelength scale}\label{sect.abs.wvl}

In order to convert measured wavelengths to energy level separations
in the ions, it is necessary to convert them to rest
wavelengths. For
RR Tel this means correctly establishing the radial velocity of the
star (or more precisely the emitting nebula). Since rest wavelengths
are accurately known for many low ionization states of elements
(typically neutral, singly and doubly-ionized stages) the basic method
employed here is to use these ions to establish the radial velocity of
the star, which can then be used to place the highly-ionized ion lines
on a rest wavelength scale. 

The standard radial
velocity used for RR Tel is the value $-62.3$~\kms\ 
originally derived by \citet{thackeray50}, although no
details were given. \citet{thackeray77} cited this earlier result and
presented new measurements that were found to be in good agreement
with the earlier measurement. Note that \citet{thackeray77} suggests
the measurements from forbidden \ion{Fe}{ii} lines are the most
reliable, and these yielded a velocity of $-61.8\pm 0.5$~\kms.

Given the high spectral resolution of both the STIS and UVES spectra,
it is appropriate to determine afresh the radial velocity using
wavelength fiducials in the spectra. Following \citet{thackeray77} the
most appropriate lines in the optical are the forbidden lines of
\ion{Fe}{ii} and these will be discussed in
Sect.~\ref{sect.uves}. The situation is more complicated at the
UV wavelengths covered by the STIS spectrum as many of the
\ion{Fe}{ii} lines are weak resonance transitions that show optical
depth effects. Sect.~\ref{sect.stis} discusses the lines chosen here
to establish a radial velocity for RR Tel.

\subsection{The STIS wavelength scale}\label{sect.stis}

This section presents the STIS emission lines that were used as
wavelength fiducials to determine the radial velocity of RR Tel and
thus allow a wavelength frame to be determined for deriving rest
wavelengths for the forbidden lines. Accurate error estimates are
important for comparing with previous measurements and so we first
give details on how these were determined.

The CALSTIS pipeline assigns 1$\sigma$ errors to the flux measurements
at each pixel in the spectra, and these were used in the Gaussian
fitting method to determine  fitting errors for each line. The second
source of error comes from the scatter in the velocities of the
wavelength fiducial lines, which we determine as the standard
deviation of the lines' velocities. The three wavelength bands were
considered separately and the error values are given later in this
section. 

A feature of echelle spectrographs is that there is often significant
wavelength overlap of adjacent spectral orders, therefore a number of
lines are measured twice. All such lines in the three wavelength bands
that were isolated
and had a good signal were selected, and their centroids determined
through Gaussian fitting. The wavelength difference between the two
occurrences of a line were determined and converted to a velocity, and
then the velocities from all the line pairs in a wavelength band
collected and the standard deviation found. For the SW, MW and LW
bands the standard deviations were found to be 2.0, 1.5 and 2.1~\kms,
respectively, and these numbers were treated as the third source of
error.

A fourth source of errors was found by considering those ions for
which an energy level has two decay paths. E.g., consider a three
level ion where level 3 decays to level 2, giving an emission line of
wavelength $\lambda_{32}$, and level 2 decays to level 1 giving an
emission line of wavelength $\lambda_{21}$. Now suppose level 3 also decays
directly to level 1 giving a line at wavelength $\lambda_{31}$. The
three wavelengths are related by $\lambda_{31}^{-1} =
\lambda_{32}^{-1} + \lambda_{21}^{-1}$, thus by measuring two of the
line wavelengths one can predict the position of the third line. Six
ions in the combined STIS and UVES  spectra allow this check to be
performed and the results and consequences on the error analysis are
discussed in Sect.~\ref{sect.consistency}. We find that this error
source dominates the others.

The rest wavelength scales for the three STIS channels were determined
independently using emission lines from low ionization species for
which laboratory wavelengths are accurately known. The criteria for
choosing these lines are as follows:
\begin{enumerate}
\item the energy levels given in the NIST database must be accurate to
  2 decimal places (or better);
\item resonance lines are not used except where optical depth effects
  are negligible; and
\item the lines are unblended and unaffected by interstellar absorption.
\end{enumerate}
Criterion 1  restricts the selection to low ionization stages
(neutral, singly-charged and doubly-charged) since the more
highly-charged ions have less accurately known energies (which is in
fact the motivation for this work). Criterion 2 follows from the fact
that most resonance lines show a significant redshift compared to
other lines, and are often asymmetric and/or broadened. For example, the strong
resonance lines \ion{N}{v} \lam1238 and \ion{C}{iv} \lam1548 yield
velocities of $-59$ and $-55$~\kms, compared to the final radial velocity
of $-65$ to $-70$~\kms\ found below. Considering lines of \ion{Fe}{ii}, the
resonance lines of UV multiplets 1 and 2 have widths of around
30--40~\kms, much larger than those of multiplets excited through
radiative pumping such as 391 and 399 which have widths of
15--25~\kms. These latter \ion{Fe}{ii} lines, although resonance
transitions themselves, are not excited from the ground levels of
\ion{Fe}{ii} and so not subject to photon trapping. They are discussed
further in Sect.~\ref{sect.fe2} below. 

The numbers of reference lines for the SW, MW and LW channels are 7, 8
and 35, respectively. The high number for the LW channel reflects the
large number of \ion{Fe}{ii} lines.  The average radial velocities
derived for the three channels are $-65.2\pm 1.8$, $-68.3\pm 1.2$ and
$-69.7\pm 1.7$~\kms. Each emission line measured in the three channels
was corrected by the relevant velocity to yield the rest wavelength.

We note that the radial velocity for the SW channel is not consistent
with those of the MW and LW channels, and all three velocities are
inconsistent with the UVES radial velocity of $-62.6\pm 1.3$~\kms\
(Sect.~\ref{sect.uves}). The latter is close to the radial velocity
found by previous authors for RR Tel, suggesting that the STIS
absolute velocity scale shows discrepancies of up to 7~\kms.
It is possible that the cool lines in the RR Tel spectrum have moved
by around 5--7~\kms\ between 1999 and 2000, but this would be
surprising given the relatively slow changes previously recorded in the RR Tel spectra.
The discrepancy between the radial velocities is not a direct problem
for the current work since we are interested in velocities \emph{relative} to
the radial velocity. The only problem would be if the low ionization,
reference lines had drifted in velocity relative to the higher
ionization lines, but the comparisons presented in
Fig.~\ref{fig.oxy.rec} suggest that this is not the case.

The following sections discuss the individual ions and emission lines used as wavelength
fiducials, and
Table~\ref{tbl.rad-vel} gives the radial velocities from the selected
reference lines.

\subsubsection{O I}

The $2s^22p^3(^4S)3s$ $^3S_1$ level in \ion{O}{i} is believed to be
excited in giant stars through radiative pumping of the
$2s^22p^3(^4S)3d$ $^3D_J$ levels by \ion{H}{i} Ly$\beta$: the $^3D$
levels decay to $2s^22p^3(^4S)3p$ $^3P_J$, which in turn decay to
the $^3S_1$ level, strongly enhancing the level's population
\citep{haisch77}. $^3S_1$ 
decays to the three $^3P_J$ levels in the ground configuration, giving
rise to lines at 1302.2, 1304.9 and 1306.0~\AA. Since \lam1302.2 is a decay to
the ground level of \ion{O}{i}, interstellar neutral oxygen strongly
absorbs the stellar spectrum at this wavelength, but the high radial
velocity of the star leaves a portion of the stellar emission line
which, however, is not useful for the present study. \lam1304.9 is
completely removed from the spectrum by interstellar \ion{Si}{ii}
\lam1304.370. The remaining line, \lam1306.0, is unaffected by interstellar
absorption.

A further decay from $^3S_1$ occurs to the $^1D_2$ level in the ground
configuration giving a line at 1641.305~\AA, and this is found in the long wavelength wing of the
very strong \ion{He}{ii} \lam1640.4. The line thus sits on a sloping
background and was fit here with a single Gaussian rather than by
attempting a double-Gaussian fit to the two lines together. The width
of \lam1641.305 is very narrow at 11.4~\kms, compared to 41.5~\kms\
for \lam1306.0. This implies that \lam1306.0 is optically thick, which is
confirmed by the flux ratio of 1.6 (\lam1641.3/\lam1306.0) which is
much larger than the theoretical branching ratio of $2\times
10^{-5}$. For this reason \lam1306.0 is not included in
Table~\ref{tbl.rad-vel}.

\subsubsection{C\,III}\label{sect.c3}

Four lines are available for \ion{C}{iii} with the strongest line,
\lam1908.7, showing a clearly asymmetric line profile with an extended
long wavelength wing. The wing is likely related to the red-shifted,
low density plasma components discussed in
Sect.~\ref{sect.cpts}. The line profile was fitted with two Gaussians,
each forced to have the same width, and the stronger, short wavelength
Gaussian was assumed to correspond to the radial velocity of the
system and the wavelength is given in Table~\ref{tbl.rad-vel}.
The weak forbidden line \lam1906.7 is found to be redshifted relative
to the expected wavelength by around 20~\kms\ and
Appendix~\ref{app.model} demonstrates that this is consistent with the
line being formed in the redshifted, low density plasma component.
It therefore is not
useful as a wavelength fiducial here.

The remaining two lines, \lam1247.4 and \lam2297.6, are significantly
weaker than \lam1908.7 but  have very similar line widths to this line
and so appear to be unblended. Both lines are listed in
Table~\ref{tbl.rad-vel}. 

The $2s2p$ $^3P_J$ -- $2p^2$ $^3P_{J^\prime}$ multiplet is found at
around 1175~\AA\ but the spectrum is noisy in this region and ratios
of the multiplet's lines vary significantly from the optically thin
case suggesting the lines are optically thick and possibly absorbed by
the interstellar medium so they are not used as velocity references.

\subsubsection{O\,III}

The intercombination lines at 1660.8 and 1666.2~\AA\ are very strong in RR
Tel and show  asymmetric profiles similar to  \ion{C}{iii} \lam1908.7
discussed in the previous section. The
centroids were derived in the same manner as the \ion{C}{iii} line
using two Gaussians of equal width. The \ion{O}{iii} lines are found
in both the SW and MW spectra but for both they are found to be around
3~\kms\ blue-shifted relative to the other wavelength reference
lines. For this reason they have not been used as reference lines.

The forbidden $^3P_1$--$^1S_0$ transition at
2321.7~\AA\ is strong in the RR Tel spectrum and has an asymmetric
profile, with the long wavelength wing being significantly stronger
than for the intercombination lines. Since the \ion{O}{iii} forbidden
lines in the visible have unusual profiles
(Sect.~\ref{sect.cpts}) it was decided not to include the
\lam2321.7 line in the present analysis.

Ten Bowen fluorescence lines of \ion{O}{iii} are found between 2800
and 3060~\AA. The strongest lines, \lam2837 and \lam3048, clearly show
asymmetric profiles similar to \lam\lam1660.8, 1666.2. The three lines at 2819.5, 3036.3
and 3060.2~\AA\ were chosen as velocity references as they are each
isolated in the spectrum and so unaffected by blending.

\subsubsection{Mg\,II}

The strong \lam2796.4, 2803.5 resonance lines are not suitable as wavelength
references as they are redshifted relative to other species,
suggesting they are affected by P Cygni like absorption on the short
wavelength side of their profiles. In addition both lines have strong
interstellar absorption features on their long wavelength sides. The
much weaker $3p$ $^2P_{3/2}$ -- $3d$ $^2D_{5/2}$ transition lies
between the two strong resonance lines at 2798.82~\AA\ and is suitable
as a wavelength reference. A further $3p$--$3d$ transition occurs at
2791.60~\AA\ but is blended, probably with a \ion{Fe}{v} transition.
$3p$ $^2P_{3/2}$ -- $4s$ $^2S_{1/2}$ is also found in the STIS
spectrum at 2937.37~\AA\ but it is very weak and is not used as a
wavelength reference here.

\subsubsection{Al\,II}\label{sect.al2}

The intercombination line at 2669.9~\AA\ is unblended and does not show
an asymmetric profile. Another \ion{Al}{ii} line is the strong
resonance transition at  1670.8~\AA, however this clearly shows
interstellar absorption on the long wavelength side of the profile and
so is not suitable as a velocity reference.

\subsubsection{Si\,II}\label{sect.si2}

The intercombination lines, $3s^23p$ $^2P_J$ -- $3s3p^2$
$^4P_{J^\prime}$, are found between 2329 and 2351~\AA. The weakest
line, \lam2329.24, is too faint to be observed while \lam2344.92 is
coincident with \ion{Fe}{ii} interstellar absorption and is not
seen. The line at 2350.89~\AA\ is close to \ion{Si}{vii} \lam2350.73,
but this line will be very weak based on the strength of \ion{Si}{vii}
\lam2147.40 and so we use \lam2350.89 as a reference.
The strongest line of the \ion{Si}{ii} multiplet is \lam2335.32
however this is notably broader than \lam2350 which is likely due to
blends from \ion{Si}{ii} \lam2335.12 (part of the same multiplet) and
\ion{Fe}{ii} \lam2335.18. It is not possible to separate these
components and so we do not use \lam2335.32 as a wavelength reference.

\subsubsection{Si\,III}\label{sect.si3}

As with other strong resonance lines, \ion{Si}{iii} \lam1206.5 is
redshifted relative to other lines in the spectrum, and it also shows
interstellar absorption on the long wavelength side of the profile. We
thus do not use it as a wavelength reference. The intercombination
line, \lam1892.0 is actually much stronger than \lam1206.5 and, like other
strong intercombination lines in the spectrum, has an asymmetric line
profile. The line has been fit with
two Gaussians of equal width and the shorter wavelength component is
taken to be at the radial velocity of the star.
\lam2542.6 is a much weaker line, but is narrow
and unblended and we list it in Table~\ref{tbl.rad-vel}.

\subsubsection{S\,III}

The $3s^23p^2$ $^3P_{1,2}$ -- $3s3p^3$ $^5S_2$ intercombination lines
at 1713.1 and 1728.9~\AA\ are narrow, although 
fairly weak, and suitable as wavelength references.

\subsubsection{Fe\,II}\label{sect.fe2}

\ion{Fe}{ii} gives rise to more lines than any other species in the RR Tel
spectrum and accurate wavelengths are known for many of them. However,
many of the \ion{Fe}{ii} lines are weak and/or blended which means
care has to be taken in selecting lines as wavelength references. In
addition it is noticeable that a number of the stronger \ion{Fe}{ii}
transitions are broad and redshifted relative to the other
transitions. Examples include $a~^6D_{7/2}$--$z~^6D_{9/2}$
(\lam2626.451),  $a~^6G_{9/2}$--$z~^6H_{11/2}$ (\lam2459.528) and
$a~^6D_{9/2}$--$z~^6F_{11/2}$ (\lam2382.765), which have velocities of
$-60$ to $-62$~\kms\ and widths of 30 to 45~\kms, compared
to $-66$ to $-70$~\kms\ and 15 to 25~\kms\ for more typical lines.

There are few \ion{Fe}{ii} lines in the STIS SW spectrum and we have
used three lines between 1360 and 1414~\AA. \lam1360.2 is a narrow,
well-observed line which was first identified as being fluoresced by
\ion{H}{i} \lya\ by \citet{johansson88b}, although \ion{O}{v}
\lam1218.3 may also contribute to the radiative pumping
\citep{hartman00}. \lam1392.1 and \lam1413.7 are both excited through
radiative pumping by \ion{He}{ii} \lam1084.9 and were identified by
\citet{hartman00} and \citet{jordan98}, respectively. 

Only two \ion{Fe}{ii} lines in the MW spectrum are used as wavelength
fiducials, and both arise from the $z~^4H_{11/2}$ level which is
radiatively pumped by \ion{C}{iv} \lam1548.2 \citep{johansson83}
giving rise to 10 lines in all that are very prominent in the RR Tel
\ion{Fe}{ii} spectrum .
Of the ten lines three 
are anomalously broad, indicating blending, and a fourth
(\lam2168.105) shows an
anomalous blueshift. The six remaining lines have
narrow widths between 20 and 25~\kms\ and velocity shifts between
$-67$ and $-70$~\kms\ and have been used as reference lines.

The remaining \ion{Fe}{ii} lines used for the wavelength calibration
are all found in the LW spectrum and 
arise from four multiplets that are excited through radiative pumping
by \ion{H}{i} Ly$\alpha$, either directly 
or by cascading from fluoresced levels. 
There are four groups of lines in all, three corresponding to 
UV multiplets 78, 391 and
399, and a fourth corresponding to the unnumbered multiplet
$z~^4P$--$e~^4D$.

UV multiplet 78 ($a~^4P$--$z~^4P$) gives rise to seven lines between
2944 and 3003~\AA, which is a region less crowded than other parts of the STIS
RR Tel spectrum. One line (\lam2965.489) is a known blend with another
\ion{Fe}{ii} transition, but the remaining six lines have narrow
widths between 20 and 23~\kms\ and velocities ranging from $-67$ to
$-71$~\kms, with an average of $-68.9$~\kms.

Five lines from UV
multiplet 391 ($z~^4F$--$e~^4D$) are found in the RR
Tel spectrum between 2840 and 2867~\AA. Each line is unblended and the
line widths range from 13 to 19~\kms, and the velocities from $-70$ to
$-71$~\kms.

UV multiplet 399 ($z~^4D$--$e~^4D$) is also emitted from the e~$^4D$
levels, and seven lines are found in the RR Tel spectrum between 2845
and 2886~\AA. One line (\lam2885.611) is significantly broader than
the others, indicating it is blended. The remaining lines have narrow
widths between $-15$ and $-18$~\kms\ and velocities between $-70$ and
$-72$~\kms. 

Six lines are observed from the unnumbered UV multiplet $z~^4P$--$e~^4D$
between 3037 and 3080~\AA. Five of these lines have never previously
been reported in the RR Tel spectrum, with the remaining line
(\lam3079.574) first being reported by \citet{jordan98}. All of the
lines appear to be unblended, with narrow 
line widths of between 16 and 25~\kms\ and velocities between $-67$
and $-73$~\kms.

\begin{deluxetable}{lllll}
\tabletypesize{\footnotesize}
\tablecaption{Wavelength fiducial lines for RR Tel STIS spectrum.\label{tbl.rad-vel}}
\tablehead{
   &
   &
   &
  \colhead{$\lambda_{\rm NIST}$} &
  \colhead{Velocity} \\
  \colhead{Channel} &
  \colhead{Ion} &
  \colhead{Transition} &
  \colhead{(\AA)} &
  \colhead{(\kms)}

}
\startdata
SW & \ion{O}{i} & $3s^23p^4$ $^1D_2$ -- $3s3p^5$ $^3S_1$ 
     & 1641.305 & $-67.0$ \\
   & \ion{C}{iii} & $2s2p$ $^1D_2$ -- $2p^2$ $^1S_0$ 
     & 1247.383 & $-65.4$ \\
   & \ion{Fe}{ii} & $b~^4D_{5/2}$ -- $(^3P)4s4p$ $^4F_{7/2}$ 
     & 1360.178 & $-67.7$ \\
   & \ion{Fe}{ii} & $b~^2H_{11/2}$ -- $u~^2G_{9/2}$ 
     & 1392.148 & $-64.0$ \\
   & \ion{Fe}{ii} & $a~^4H_{11/2}$ -- $sp~x~^4H_{11/2}$ 
     & 1413.702 & $-62.8$ \\
   & \ion{S}{iii} & $3s^23p^2$ $^3P_1$ -- $3s3p^3$ $^5S_2$ 
     & 1713.114 & $-63.7$ \\
   & \ion{S}{iii} & $3s^23p^2$ $^3P_2$ -- $3s3p^3$ $^5S_2$ 
     & 1728.942 & $-65.7$ \\
\noalign{\medskip}
MW & \ion{S}{iii} & $3s^23p^2$ $^3P_1$ -- $3s3p^3$ $^5S_2$ 
     & 1713.114 & $-69.0$ \\
   & \ion{S}{iii} & $3s^23p^2$ $^3P_2$ -- $3s3p^3$ $^5S_2$ 
     & 1728.942 & $-69.2$ \\
   & \ion{Si}{iii} & $3s^2$ $^1S_0$ -- $3s3p$ $^3P_1$ 
     & 1892.030 & $-66.6$ \\
   & \ion{C}{iii} & $2s^2$ $^1S_0$ -- $2s2p$ $^3P_1$
     & 1908.734 & $-69.1$ \\
   & \ion{Fe}{ii} & $a~^4H_{13/2}$ -- $y~^4H_{11/2}$
     & 2211.806 & $-70.1$ \\
   & \ion{Fe}{ii} & $a~^4H_{11/2}$ -- $y~^4H_{11/2}$
     & 2220.585 & $-67.2$ \\
   & \ion{C}{iii} & $2s2p$ $^1P_1$ -- $2p^2$ $^1D_2$
     & 2297.587 & $-67.6$ \\
   & \ion{Si}{ii} & $3s^23p$ $^2P_{3/2}$ -- $3s3p^2$ $^4P_{1/2}$
     & 2350.892 & $-66.6$ \\
\noalign{\medskip}
LW & \ion{C}{iii} & $2s2p$ $^1P_1$ -- $2p^2$ $^1D_2$
     & 2297.587 & $-69.8$ \\
   & \ion{Si}{ii} & $3s^23p$ $^2P_{3/2}$ -- $3s3p^2$ $^4P_{1/2}$
     & 2350.892 & $-66.7$ \\
   & \ion{Fe}{ii} & $a~^4G_{11/2}$ -- $y~^4H_{11/2}$
     & 2436.959 & $-68.9$ \\
   & \ion{Fe}{ii} & $b~^2H_{11/2}$ -- $y~^4H_{11/2}$
     & 2481.799 & $-69.1$ \\
   & \ion{Fe}{ii} & $b~^2H_{9/2}$ -- $y~^4H_{11/2}$
     & 2493.096 & $-66.6$ \\
   & \ion{Si}{iii} & $3s3p$ $^1P_1$ -- $3p^2$ $^1D_2$ 
     & 2542.581 & $-66.6$ \\
   & \ion{Al}{ii} & $3s^2$ $^1S_0$ -- $3s3p$ $^3P_1$ 
     & 2669.948 & $-69.7$ \\
   & \ion{Fe}{ii} & $b~^2G_{9/2}$ -- $y~^4H_{11/2}$
     & 2772.004 & $-68.3$ \\
   & \ion{Mg}{ii} & $3p$ $^2P_{3/2}$ -- $3d$ $^2D_{5/2}$
     & 2798.823 & $-70.3$ \\
   & \ion{O}{iii} & $2p3p$ $^3D_2$ -- $2p3d$ $^3P_2$
     & 2819.527 & $-70.4$ \\
   & \ion{Fe}{ii} & $z~^4F_{9/2}$ -- $e~^4D_{7/2}$ 
     & 2840.348 & $-69.6$ \\
   & \ion{Fe}{ii} & $z~^4D_{1/2}$ -- $e~^4D_{1/2}$ 
     & 2845.795 & $-70.1 $ \\
   & \ion{Fe}{ii} & $z~^4D_{3/2}$ -- $e~^4D_{3/2}$ 
     & 2846.261 & $-71.1$ \\
   & \ion{Fe}{ii} & $z~^4F_{7/2}$ -- $e~^4D_{5/2}$ 
     & 2846.433 & $-70.1$ \\
   & \ion{Fe}{ii} & $z~^4D_{5/2}$ -- $e~^4D_{5/2}$ 
     & 2848.944 & $-70.7$ \\
   & \ion{Fe}{ii} & $z~^4F_{5/2}$ -- $e~^4D_{3/2}$ 
     & 2849.157 & $-70.7$ \\
   & \ion{Fe}{ii} & $z~^4F_{3/2}$ -- $e~^4D_{1/2}$ 
     & 2852.561 & $-69.4$ \\
   & \ion{Fe}{ii} & $z~^4D_{7/2}$ -- $e~^4D_{7/2}$ 
     & 2857.748 & $-70.4$ \\
   & \ion{Fe}{ii} & $z~^4D_{1/2}$ -- $e~^4D_{3/2}$ 
     & 2859.469 & $-71.4$ \\
   & \ion{Fe}{ii} & $z~^4F_{1/2}$ -- $e~^4D_{1/2}$ 
     & 2866.301 & $-71.4$ \\
   & \ion{Fe}{ii} & $z~^4D_{3/2}$ -- $e~^4D_{5/2}$ 
     & 2870.155 & $-70.7$ \\
   & \ion{Fe}{ii} & $a~^4P_{1/2}$ -- $z~^4P_{3/2}$ 
     & 2945.257 & $-70.5$ \\
   & \ion{Fe}{ii} & $a~^4P_{5/2}$ -- $z~^4P_{3/2}$ 
     & 2948.516 & $-68.6$ \\
   & \ion{Fe}{ii} & $a~^4P_{3/2}$ -- $z~^4P_{3/2}$ 
     & 2965.899 & $-69.2$ \\ 
   & \ion{Fe}{ii} & $a~^4P_{5/2}$ -- $z~^4P_{5/2}$ 
     & 2985.695 & $-67.0$ \\
   & \ion{Fe}{ii} & $a~^4P_{3/2}$ -- $z~^4P_{1/2}$ 
     & 2986.416 & $-69.0$ \\
   & \ion{Fe}{ii} & $a~^4P_{3/2}$ -- $z~^4P_{5/2}$ 
     & 3003.521 & $-69.2$ \\
   & \ion{O}{iii} & $2p3s$ $^3P_1$ -- $2p3p$ $^3P_1$
     & 3036.298 & $-69.0$ \\
   & \ion{Fe}{ii} & $z~^4P_{5/2}$ -- $e~^4D_{5/2}$ 
     & 3037.847 & $-71.1$ \\
   & \ion{Fe}{ii} & $z~^4P_{3/2}$ -- $e~^4D_{3/2}$ 
     & 3049.878 & $-73.1$ \\
   & \ion{Fe}{ii} & $z~^4P_{1/2}$ -- $e~^4D_{1/2}$ 
     & 3056.240 & $-70.8$ \\
   & \ion{O}{iii} & $2p3s$ $^3P_2$ -- $2p3p$ $^3P_1$
     & 3060.199 & $-74.1$ \\
   & \ion{Fe}{ii} & $z~^4P_{1/2}$ -- $e~^4D_{3/2}$ 
     & 3072.017 & $-69.5$ \\
   & \ion{Fe}{ii} & $z~^4P_{3/2}$ -- $e~^4D_{5/2}$ 
     & 3077.329 & $-70.0$ \\
   & \ion{Fe}{ii} & $z~^4P_{5/2}$ -- $e~^4D_{7/2}$ 
     & 3079.574 & $-67.5$ \\
\enddata
\end{deluxetable}

\subsection{The UVES wavelength scale}\label{sect.uves}

For the UVES spectra, 23 emission lines of \ion{Fe}{ii} were used to
derive the radial velocity of the star which was then subtracted  to
yield the absolute wavelength scale. The emission lines are
principally forbidden lines, although the strong allowed multiplet,
a~$^4P$--z~$^4D$, was also used. The full list of transitions with
measured and rest wavelengths, and derived velocities are shown
in Table~\ref{tbl.uves.fe2}. Rest wavelengths have been derived using
the \ion{Fe}{ii} experimental energies tabulated by
\citet{fuhr06}. The average velocity is $-62.6$~\kms, 
with a standard deviation of 1.3~\kms. By comparing measured centroids
of lines observed in two spectral orders we estimate individual
centroid measurements are accurate to approximately
$\pm$~1.5~\kms. Combining these two uncertainties with that of the
wavelength consistency check discussed in Sect.~\ref{sect.consistency}
yields the final error estimate for line centroids measured from the
UVES spectra.
We note that the radial velocity derived from the UVES \ion{Fe}{ii}
lines is in good agreement with the values of \citet{thackeray50} and
\citet{thackeray77}.

\begin{deluxetable}{lrll}
\tablecaption{UVES \ion{Fe}{ii} velocities.\label{tbl.uves.fe2}}
\tablehead{
  \colhead{$\lambda_{\rm vac}$} &
  &
  &\colhead{Velocity} \\
  \colhead{(\AA)} &
  \colhead{Multiplet} &
  \colhead{Transition} &
  \colhead{(\kms)} 
}
\startdata
  3187.659 &  45 &                     a~$^4P_{3/2}$ -- z~$^4D_{3/2}$ & $ -63.9$ \\
  3193.832 &  45 &                     a~$^4P_{5/2}$ -- z~$^4D_{5/2}$ & $ -64.1$ \\
  3194.722 &  45 &                     a~$^4P_{1/2}$ -- z~$^4D_{1/2}$ & $ -64.2$ \\
  3211.371 &  45 &                     a~$^4P_{1/2}$ -- z~$^4D_{3/2}$ & $ -62.7$ \\
  3214.237 &  45 &                     a~$^4P_{3/2}$ -- z~$^4D_{5/2}$ & $ -63.0$ \\
  3228.674 &  45 &                     a~$^4P_{5/2}$ -- z~$^4D_{7/2}$ & $ -63.2$ \\
  4815.880 &  21 &                     a~$^4F_{9/2}$ -- b~$^4F_{9/2}$ & $ -64.7$ \\
  4890.982 &   5 &                     a~$^6D_{7/2}$ -- b~$^4P_{5/2}$ & $ -64.1$ \\
  4906.709 &  21 &                     a~$^4F_{7/2}$ -- b~$^4F_{7/2}$ & $ -61.8$ \\
  5109.365 &  19 &                     a~$^4F_{5/2}$ -- b~$^4P_{1/2}$ & $ -60.1$ \\
  5113.051 &  20 &                    a~$^4F_{9/2}$ -- a~$^4H_{11/2}$ & $ -61.8$ \\
  5165.390 &  37 &                     a~$^4D_{7/2}$ -- a~$^2F_{7/2}$ & $ -64.7$ \\
  5183.391 &  19 &                     a~$^4F_{3/2}$ -- b~$^4P_{1/2}$ & $ -62.2$ \\
  5221.512 &  20 &                     a~$^4F_{7/2}$ -- a~$^4H_{9/2}$ & $ -62.1$ \\
  5263.085 &  20 &                    a~$^4F_{7/2}$ -- a~$^4H_{11/2}$ & $ -61.6$ \\
  5270.341 &  19 &                     a~$^4F_{5/2}$ -- b~$^4P_{3/2}$ & $ -61.5$ \\
  5274.814 &  19 &                     a~$^4F_{9/2}$ -- b~$^4P_{5/2}$ & $ -61.4$ \\
  5298.303 &  20 &                     a~$^4F_{5/2}$ -- a~$^4H_{7/2}$ & $ -61.2$ \\
  5335.129 &  20 &                     a~$^4F_{5/2}$ -- a~$^4H_{9/2}$ & $ -63.0$ \\
  5377.947 &  20 &                     a~$^4F_{3/2}$ -- a~$^4H_{7/2}$ & $ -62.9$ \\
  5434.640 &  19 &                     a~$^4F_{7/2}$ -- b~$^4P_{5/2}$ & $ -63.0$ \\
  5478.764 &  36 &                     a~$^4D_{3/2}$ -- b~$^2P_{1/2}$ & $ -59.9$ \\
  5748.560 &  36 &                     a~$^4D_{5/2}$ -- b~$^2P_{3/2}$ & $ -63.2$ \\

\enddata
\end{deluxetable}

\begin{deluxetable}{lllllll}
\tabletypesize{\footnotesize}
\tablecaption{Measured wavelengths.\label{tbl.wavelengths}}
\tablehead{
  \colhead{Transition} &
  \colhead{$\lambda_{\rm vac}$ (\AA)} &
  \colhead{$\lambda_{\rm air}$ (\AA)} &
  \colhead{$\sigma_\lambda$ (\AA)} &
  \colhead{$\lambda_{\rm NIST}$}\tablenotemark{a} (\AA) &
  \colhead{Previous (\AA)} &
  \colhead{Source\tablenotemark{b}} 
}
\startdata
\sidehead{Beryllium isoelectronic sequence}
\sidehead{\ion{N}{iv}}
$2s^2$ $^1S_0$ -- $2s2p$ $^3P_1$ &
   1486.502 & 1486.502 & 0.030 & 1486.496 & $1486.48\pm 0.05$ & P83 \\
   &&&&& $1486.51\pm 0.04$ & D76 \\
   &&&&& $1486.52\pm 0.02$ & S77 \\
\sidehead{\ion{O}{v}}
$2s^2$ $^1S_0$ -- $2s2p$ $^3P_2$ &
   1213.807 & 1213.807 & 0.029 & 1213.809 & $1213.90\pm 0.05$ & S77 \\
$2s^2$ $^1S_0$ -- $2s2p$ $^3P_1$ &
   1218.349 & 1218.349 & 0.025 & 1218.344 & $1218.35\pm 0.02$ & S77\\
   &&&&& $1218.35\pm 0.04$ & D76 \\
   &&&&& $1218.37\pm 0.05$ & P83\\
\sidehead{Boron isoelectronic sequence}
\sidehead{\ion{C}{ii}}
$^2P_{1/2}$--$^4P_{3/2}$ &
   2324.272 & 2323.558 & 0.047 & 2324.214 & $2323.42\pm 0.05$ & P83
   (air)\\
   &&&&&$2323.52\pm 0.04$ & D77 (air) \\
$^2P_{1/2}$--$^4P_{1/2}$ &
   2325.411 & 2324.697 & 0.046  & 2325.403 & $2324.69\pm 0.05$ &P83 (air)\\
   &&&&&$2324.72\pm 0.04$ & D77 (air) \\
$^2P_{3/2}$--$^4P_{5/2}$ &
   2326.120 & 2325.406 & 0.045 & 2326.113 & $2325.38\pm 0.05$ &P83 (air)\\
   &&&&&$2325.40\pm 0.04$ & D77 (air)\\
$^2P_{3/2}$--$^4P_{3/2}$ &
   2327.669 & 2326.954 & 0.047 & 2327.645 & $2326.92\pm 0.05$ &P83 (air)\\
   &&&&&$2326.98\pm 0.04$ & D77 (air)\\
$^2P_{3/2}$--$^4P_{1/2}$ &
   2328.871 & 2328.156 & 0.046 & 2328.838 & $2328.08\pm 0.05$ &P83 (air)\\
   &&&&&$2328.14\pm 0.04$ & D77 (air) \\
\sidehead{\ion{N}{iii}}
$^2P_{1/2}$--$^4P_{3/2}$ &
   1746.816 & 1746.816 & 0.037 & 1746.823 & $1746.81\pm 0.02$ & S77\\
   &&&&& $1746.82\pm 0.04$ & D76 \\
   &&&&& $1746.77\pm 0.05$ & P83 \\
$^2P_{1/2}$--$^4P_{1/2}$ &
   1748.637 & 1748.637 & 0.035 & 1748.646 & $1748.63\pm 0.02$ & S77\\
   &&&&& $1748.63\pm 0.04$ & D76 \\
   &&&&& $1748.63\pm 0.05$ & P83 \\
$^2P_{3/2}$--$^4P_{5/2}$ &
   1749.663 & 1749.663 & 0.034 & 1749.674 & $1749.67\pm 0.01$ & S77\\
   &&&&& $1749.67\pm 0.04$ & D76 \\
   &&&&& $1749.64\pm 0.05$ & P83 \\
$^2P_{3/2}$--$^4P_{3/2}$ &
   1752.139 & 1752.139 & 0.035 & 1752.160 & $1752.14\pm 0.01$ & S77\\
   &&&&& $1752.12\pm 0.04$ & D76 \\
   &&&&& $1752.11\pm 0.05$ & P83 \\
$^2P_{3/2}$--$^4P_{1/2}$ &
   1753.974 & 1753.974 & 0.035 & 1753.995 & $1753.98\pm 0.01$ & S77\\
   &&&&& $1753.98\pm 0.04$ & D76 \\
   &&&&& $1753.96\pm 0.05$ & P83 \\
\sidehead{\ion{O}{iv}}
$^2P_{1/2}$--$^4P_{3/2}$ &
   1397.199 & 1397.199 & 0.029 & 1397.232 & $1397.20\pm 0.04$ & D76 \\
   &&&&& $1397.22\pm 0.02$ & S77\\
   &&&&& $1397.18\pm 0.05$ & P83 \\
   &&&&& $1397.219\pm 0.075$ & H99 \\
   &&&&& $1397.166\pm 0.004$ & K02 \\
$^2P_{1/2}$--$^4P_{1/2}$ &
   1399.766 & 1399.766 & 0.029 & 1399.780 & $1399.77\pm 0.04$ & D76 \\
   &&&&& $1399.78\pm 0.01$ & S77\\
   &&&&& $1399.75\pm 0.05$ & P83 \\
   &&&&& $1399.785\pm 0.075$ & H99 \\
   &&&&& $1399.731\pm 0.004$ & K02\\
$^2P_{3/2}$--$^4P_{5/2}$ &
   1401.157 & 1401.157 & 0.029 & 1401.157 & $1401.16\pm 0.04$ & D76 \\
   &&&&& $1401.17\pm 0.01$ & S77\\
   &&&&& $1401.14\pm 0.05$ & P83 \\
   &&&&& $1401.168\pm 0.075$ & H99\\
   &&&&& $1401.115\pm 0.004$ & K02\\
$^2P_{3/2}$--$^4P_{3/2}$ &
   1404.783 & 1404.783 & 0.029 & 1404.806 & $1404.79\pm 0.04$ & D76 \\
   &&&&& $1404.80\pm 0.02$ & S77\\
   &&&&& $1404.77\pm 0.05$ & P83 \\
   &&&&& $1404.797\pm 0.075$ & H99\\
   &&&&& $1404.740\pm 0.004$ & K02\\
$^2P_{3/2}$--$^4P_{1/2}$ &
   1407.372 & 1407.372 & 0.029 & 1407.382 & $1407.38\pm 0.04$ & D76 \\
   &&&&& $1407.39\pm 0.02$ & S77\\
   &&&&& $1407.36\pm 0.05$ & P83 \\
   &&&&& $1407.387\pm 0.075$ & H99\\
   &&&&& $1407.333\pm 0.004$ & K02 \\
\sidehead{Carbon isoelectronic sequence}
\sidehead{\ion{N}{ii}}
$^1D_2$--$^1S_0$ &
   5756.205 & 5754.607 & 0.116 & 5756.191 & $5754.57\pm 0.04$ & B60 (air)\\
$^3P_1$--$^1S_0$ &
   3063.791 & 3062.900 & 0.129 & 3063.716 & $3062.82\pm 0.02$ & B60 (air)\\
$^3P_1$--$^5S_2$ &
   2139.683 & 2139.009 & 0.045 & 2139.683 & $2138.88\pm 0.05$ & P83 (air)\\
\sidehead{\ion{Ne}{v}}
$^3P_1$--$^1D_2$ &
   3346.820 & 3345.858 & 0.065 & 3346.783 & $3345.83 \pm 0.02$ & B60 (air)\\
$^3P_2$--$^1D_2$ &
   3426.905 & 3425.923 & 0.067 & 3426.864 & $3425.87 \pm 0.02$ & B60 (air)\\
$^3P_1$--$^1S_0$ &
   1574.671 & 1574.671 & 0.032 & 1574.700 & $1575.2\pm 1.0$ & B60 \\
   &&&&& $1574.68\pm 0.05$ & P83 \\
$^3P_2$--$^1S_0$ &
   1592.187 & 1592.187 & 0.045 & 1592.206 & $1592.7\pm 1.0$ & B60 \\
$^1D_2$--$^1S_0$ &
   2973.968 & 2973.101 & 0.061 & 2974.002 & $2974.8\pm 3.0$ & B60
   (air) \\
   &&&&& $2974.00 \pm 0.05$ & P83
   (vac) \\
$^3P_2$--$^5S_2$ &
   1145.591 & 1145.591 & 0.025 & 1145.606 & $1145.61\pm 0.02$ & S77 \\
   &&&&& $1145.62\pm 0.02$ & Y05 \\
\sidehead{\ion{Na}{vi}}
$^3P_1$--$^1D_2$ &
   2872.650 & 2871.808 & 0.060 & 2873.563 & $2871.1\pm 3.0$ & B60 (air)\\
$^3P_2$--$^1D_2$ &
   2971.785 & 2970.918 & 0.061 & 2972.740 & $2970.0\pm 3.0$ & B60 (air)\\
$^1D_2$--$^1S_0$ &
   2569.588 & 2568.818 & 0.097 & 2569.637 & $2568.9\pm 3.0$ & B60 (air)\\
$^3P_1$--$^1S_0$ &
   1356.321 & 1356.321 & 0.039 & 1356.558 & $1356.2\pm 1.0$ & B60 \\
\sidehead{Nitrogen isoelectronic sequence}
\sidehead{\ion{O}{ii}}
$^4S_{3/2}$--$^2P_{3/2}$ &
   2471.200\tablenotemark{c} & 2470.453 & 0.051 & 2471.088 & $2470.30 \pm 0.02$ & B60
   (air) \\
\sidehead{\ion{Ne}{iv}}
$^4S_{3/2}$--$^2P_{3/2}$ &
   1601.502 & 1601.502 & 0.033 & 1601.504 & $1602.0\pm 3.0$ & B60 \\
   &&&&& $1601.47\pm 0.05$ & P83 \\
$^4S_{3/2}$--$^2P_{1/2}$ &
   1601.698 & 1601.698 & 0.034 & 1601.676 & $1602.1\pm 3.0$ & B60 \\
$^4S_{3/2}$--$^2D_{3/2}$ &
   2422.617\tablenotemark{c} & 2421.881 & 0.050 & 2422.510 & $2422.8\pm 3.0$ & B60 (air)\\
   &&&&& $2422.43 \pm 0.05$ & P83 (vac)\\
   &&&&& $2421.825\pm 0.011$ & J98 (air)\\
$^4S_{3/2}$--$^2D_{5/2}$ &
   2425.212\tablenotemark{c} & 2424.475 & 0.054 & 2425.148 & $2425.4\pm 3.0$ & B60 (air)\\
   &&&&& $2424.97 \pm 0.05$ & P83 (vac)\\
   &&&&& $2424.403\pm 0.013$ & J98 (air) \\
\sidehead{\ion{Na}{v}}
$^4S_{3/2}$--$^2D_{5/2}$ &
   2069.919\tablenotemark{c} & 2069.258 & 0.062 & 2069.108 & \nodata & \nodata \\
$^4S_{3/2}$--$^2P_{3/2}$ &
   1365.388 & 1365.388 & 0.028 & 1365.095 & \nodata & \nodata \\
$^4S_{3/2}$--$^2P_{1/2}$ &
   1366.081 & 1366.081 & 0.029 & 1365.784 & \nodata & \nodata \\
\sidehead{\ion{Mg}{vi}}
$^2D_{3/2}$--$^2P_{1/2}$ &
   3503.182 & 3502.181 & 0.068 & 3502.971 & $3503.0 \pm 3.0$ & B60 (air) \\
$^2D_{3/2}$--$^2P_{3/2}$ &
   3489.892 & 3488.893 & 0.068 & 3489.720 & $3488.1 \pm 3.0$ & B60 (air) \\
$^2D_{5/2}$--$^2P_{3/2}$ &
   3488.100 & 3487.102 & 0.068 & 3487.675 & $3485.5\pm 3.0$ & B60 (air) \\
$^4S_{3/2}$--$^2P_{3/2}$ &
   1190.040 & 1190.040 & 0.024 & 1190.074 & $1190.07\pm 0.01$ & S77 \\
   &&&&& $1190.09\pm 0.03$ & C04 \\
$^4S_{3/2}$--$^2P_{1/2}$ &
   1191.588 & 1191.588 & 0.024 & 1191.611 & $1191.62\pm 0.02$ & S77 \\
   &&&&& $1191.64\pm 0.03$ & C04 \\
$^4S_{3/2}$--$^2D_{3/2}$ &
   1805.882 & 1805.882 & 0.035 & 1805.941 & $1805.94\pm 0.03$ & S77 \\
\sidehead{Oxygen isoelectronic sequence}
\sidehead{\ion{Ne}{iii}}
$^3P_1$--$^1S_0$ &
   1814.645 & 1814.645 & 0.037 & 1814.559 & $1814.65 \pm 0.01$ & B60 \\
$^3P_2$--$^1D_2$ &
   3869.849 & 3868.752 & 0.078 & 3869.861 & $3868.76 \pm 0.02$ & B60 (air)\\
$^1D_2$--$^1S_0$ &
   3343.414 & 3342.453 & 0.067 & 3343.142 & $3342.5 \pm 0.3$ & B60 (air) \\
\sidehead{\ion{Na}{iv}}
$^3P_2$--$^1D_2$ &
   3242.660 & 3241.725 & 0.063 & 3242.563 & $3241.68 \pm 0.10$ & B60 (air)\\
$^3P_1$--$^1D_2$ &
   3363.260 & 3362.294 & 0.068 & 3363.210 & $3362.20 \pm 0.10$ & B60 (air) \\
\sidehead{\ion{Mg}{v}}
$^3P_1$--$^1S_0$ &
   1324.435 & 1324.435 & 0.027 & 1324.575 & $1324.4\pm 1.0$ & B60 \\
   &&&&& $1324.44\pm 0.01$ & S77 \\
   &&&&& $1324.40\pm 0.05$ & P83 \\
$^1D_2$--$^1S_0$ &
   2417.628 & 2416.893 & 0.052 & 2418.204 & $2416.8\pm 3.0$ & B60 (air)\\
$^3P_2$--$^1D_2$ &
   2783.644 & 2782.823 & 0.057 & 2783.499 & $2783.1\pm 3.0$ & B60 (air)\\
   &&&&& $2783.66\pm 0.05$ & P83 (vac)\\
$^3P_1$--$^1D_2$ &
   2928.991 & 2928.135 & 0.060 & 2928.867 & $2928.3\pm 3.0$ & B60 (air)\\
   &&&&& $2928.99\pm 0.05$ & P83 (vac)\\
\sidehead{\ion{Al}{vi}}
$^3P_2$--$^1D_2$ &
   2430.248 & 2429.511 & 0.050 & 2429.130 & $2430.0\pm 10.0$ & B60 (air)\\
   &&&&& 2429.499 & J98 (air)\\
$^3P_1$--$^1D_2$ &
   2603.123 & 2602.345 & 0.056 & 2601.795 & $2603.0\pm 10.0$ & B60 (air)\\
\sidehead{Magnesium isoelectronic sequence}
\sidehead{\ion{P}{iv}}
$3s^2$ $^1S_0$ -- $3s3p$ $^3P_1$ &
   1467.434 & 1467.434 & 0.031 & 1467.427 & $1467.44\pm 0.03$ & S77 \\
   &&&&& $1467.41\pm 0.05$ & P83\\
\sidehead{\ion{S}{v}}
$3s^2$ $^1S_0$ -- $3s3p$ $^3P_1$ &
   1199.162 & 1199.162 & 0.025 & 1199.134 & $1199.18\pm 0.01$ & S77\\
   &&&&& $1199.14\pm 0.05$ & P83\\
\sidehead{Aluminium isoelectronic sequence}
\sidehead{\ion{S}{iv}}
$^2P_{1/2}$--$^4P_{3/2}$ &
   1398.065 & 1398.065 & 0.041 & 1398.040 & $1397.98\pm 0.05$ & P83 \\
   &&&&&$1398.044 \pm 0.004$ & K02\\
$^2P_{3/2}$--$^4P_{5/2}$ &
   1406.043 & 1406.043 & 0.029 & 1406.016 & $1406.03\pm 0.05$ & P83 \\
   &&&&&$1406.06\pm 0.01$ & S77 \\
   &&&&&$1406.052\pm 0.075$ & H99 \\
   &&&&&$1406.004\pm 0.004$ & K02\\
$^2P_{3/2}$--$^4P_{3/2}$ &
   1416.912 & 1416.912 & 0.029 & 1416.887 & $1416.91\pm 0.05$ & P83 \\
   &&&&&$1416.93\pm 0.01$ & S77 \\
   &&&&&$1416.925\pm 0.075$ & H99 \\
   &&&&&$1416.872\pm 0.004$ & K02 \\
$^2P_{3/2}$--$^4P_{1/2}$ &
   1423.857 & 1423.857 & 0.030 & 1423.839 & $1423.89\pm 0.01$ & S77\\
   &&&&&$1423.860\pm 0.075$ & H99 \\
   &&&&&$1423.790\pm 0.004$ & K02 \\
\sidehead{Silicon isoelectronic sequence}
\sidehead{\ion{Cl}{iv}}
$^3P_1$--$^1S_0$ & 
   3119.549 & 3118.645 & 0.063 & 3119.560 & $3118.66 \pm 0.04$ & B60
   (air) \\
$^1D_2$--$^1S_0$ &
   5324.697 & 5323.214 & 0.107 & 5324.757 & $5323.29 \pm 0.10$ & B60 (air)\\
\sidehead{\ion{Ar}{v}}
$^3P_1$--$^1D_2$ &
   6436.946 & 6435.165 & 0.130 & 6437.629 & $6435.10\pm 0.10$ & B60 (air)\\
    &&&&& 6435.10 & T77 (air)\\
$^3P_1$--$^1S_0$ & 
   2691.848 & 2691.049 & 0.056 & 2692.024 & $2691.09\pm 0.04$ & B60 (air)\\
    &&&&& $2691.74\pm 0.05$ & P83 (vac) \\
\sidehead{\ion{K}{vi}}
$^3P_1$--$^1D_2$ & 
   5603.816 & 5602.259 & 0.113 & 5603.999 & $5603.2 \pm 3.0$ & B60 (air)\\
$^3P_2$--$^1D_2$ & 
   6230.117 & 6228.391 & 0.122 & 6230.297 & $6229.2 \pm 3.0$ &
B60 (air)\\
$^3P_1$--$^1S_0$ & 
   2368.265 & 2367.541 & 0.126 & 2368.243 & $2366.8\pm 3.0$ & B60 (air)\\
\sidehead{\ion{Ca}{vii}}
$^3P_1$--$^1D_2$ & 
   4940.649 & 4939.270 & 0.096 & 4940.931 & $4939.0 \pm 10.0$ &
B60 (air)\\
    &&&&& $4939.48\pm 0.20$ & T74 (air)\\
$^3P_2$--$^1D_2$ & 
   5620.140 & 5618.579 & 0.110 & 5620.314 & $5616.0 \pm 10.0$ & B60 (air)\\
    &&&&& $5618.58\pm 0.20$ & T74 (air)\\
$^3P_1$--$^1S_0$ & 
   2111.488 & 2110.819 & 0.041 & 2111.643 & \nodata & \nodata \\
\sidehead{Phosphorus isoelectronic sequence}
\sidehead{\ion{Ar}{iv}}
$^4S_{3/2}$--$^2P_{3/2}$ & 
    2854.583 & 2853.744 & 0.059 & 2854.484 & $2853.64\pm 0.04$ & B60 (air)\\
    &&&&& $2853.67\pm 0.05$ & P83 (air) \\
$^4S_{3/2}$--$^2P_{1/2}$ & 
    2869.087 & 2868.245 & 0.059 & 2868.988 & $2868.16\pm 0.10$ & B60 (air)\\
    &&&&& $2868.21\pm 0.05$ & P83 (air) \\
\sidehead{\ion{K}{v}}
$^4S_{3/2}$--$^2P_{3/2}$ & 
    2494.968 & 2494.215 & 0.065 & 2494.998 & $2494.5\pm 1.0$ & B60 (air)\\
$^4S_{3/2}$--$^2P_{1/2}$ & 
    2515.351 & 2514.594 & 0.141 & 2515.211 & $2514.5\pm 1.0$ & B60 (air)\\
$^2D_{3/2}$--$^2P_{3/2}$ & 
    6223.679 & 6221.956 & 0.125 & 6223.666 & $6223.0\pm 10.0$ & B60 (air)\\
\sidehead{\ion{Ca}{vi}}
$^4S_{3/2}$--$^2P_{3/2}$ & 
    2215.177 & 2214.486 & 0.057 & 2215.198 & $2215.0\pm 1.0$ & B60 (air)\\
    &&&&& $2215.15 \pm 0.05$ & P83 (vac)\\
$^4S_{3/2}$--$^2P_{1/2}$ & 
    2242.704 & 2242.008 & 0.044 & 2242.821 & $2242.6\pm 1.0$ & B60
    (air) \\
    &&&&& $2242.69 \pm 0.05$ & P83 (vac)\\
$^4S_{3/2}$--$^2D_{3/2}$ & 
    3726.359 & 3725.300 & 0.075 & 3726.463 & $3727.1\pm 3.0$ & B60 (air)\\
$^2D_{3/2}$--$^2P_{1/2}$ &
    5632.941 & 5631.376 & 0.110 & 5633.295 & $5631.0 \pm 10.0$ & B60 (air) \\
$^2D_{3/2}$--$^2P_{3/2}$ &
    5462.193 & 5460.674 & 0.107 & 5462.213 & $5460.0 \pm 10.0$ & B60 (air) \\
    &&&&& 5460.7 & T77 (air) \\
$^2D_{5/2}$--$^2P_{3/2}$ &
    5587.766 & 5586.213 & 0.109 & 5587.810 & $5587.0 \pm 10.0$ & B60 (air) \\
    &&&&& 5586.2 & T77 (air)\\
\sidehead{Sulphur isoelectronic sequence}
\sidehead{\ion{Ar}{iii}}
$^3P_1$--$^1S_0$ &
    3110.065 & 3109.163 & 0.072 & 3110.077 & $3109.16\pm 0.04$ & B60 (air)\\
    &&&&& 3109.04 & T77 (air)\\
    &&&&& $3109.01\pm 0.05$ & P83 (air) \\
$^1D_2$--$^1S_0$ &
    5193.141 & 5191.694 & 0.105 & 5193.262 & $5191.82 \pm 0.10$ & B60 (air)\\
    &&&&& 5191.65 & T77 (air)\\
\sidehead{\ion{K}{iv}}
$^3P_2$--$^1D_2$ &
    6103.525 & 6101.834 & 0.123 & 6103.479 & $6101.83 \pm 0.10$ & B60 (air)\\
    &&&&& 6101.76 & T77 (air) \\
\sidehead{\ion{Ca}{v}}
$^3P_1$--$^1S_0$ &
    2413.600 & 2412.866 & 0.050 & 2413.605 & $2412.4\pm 1.0$ & B60 (air)\\
    &&&&& $2413.57\pm 0.05$ & P83 (vac)\\
$^3P_2$--$^1D_2$ &
    5310.792 & 5309.313 & 0.104 & 5310.590 & $5309.18 \pm 0.10$ & B60 (air)\\
    &&&&& 5309.26 & T77 (air) \\
\enddata
\tablenotetext{a}{Vacuum wavelengths derived from energy levels available in
  version~3 of the NIST database.}
\tablenotetext{b}{References for previous wavelength
  measurements. Codes are: B60 -- \citet{bowen60}; T74 --
  \citet{thackeray74}; D76 -- \citet{doschek76}; D77 --
  \citet{doschek77}; S77 -- \citet{sandlin77}; T77 -- 
  \citet{thackeray77}; P83 -- \citet{penston83}; J98 --
  \citet{jordan98}; H99 -- \citet{harper99}; K02 -- \citet{keenan02}; C04 --
  \citet{curdt04}; Y05 -- \citet{young05b}. For lines above 2000~\AA,
  air or vacuum wavelengths are indicated.}
\tablenotetext{c}{Wavelength may be affected by the low density plasma
  component of the nebula.}
\end{deluxetable}

\subsection{Consistency checks}\label{sect.consistency}

For some of the ions considered in the present work, levels can decay
by multiple routes to the ground term of the ion and so if all lines in the
decay routes can be measured they will serve as a check on the
wavelength scales employed for the STIS and UVES spectra. One example
is the $^1S_0$ level in carbon, oxygen, silicon and sulphur-like
ions which can decay directly to $^3P_1$, or to $^3P_1$ via the
$^1D_2$ level. For nitrogen and phosphorus-like ions the
$^2P_{1/2,3/2}$ levels can decay directly to the ground $^4S_{3/2}$
level, or via the $^2D_{3/2,5/2}$ levels. 

There are six ions for which these consistency checks can be
performed in the present work: \ion{Ne}{v}, \ion{Na}{vi},
\ion{Mg}{vi}, \ion{Ne}{iii}, \ion{Mg}{v} and \ion{Ca}{vi}. Three ions,
\ion{Ne}{v}, \ion{Mg}{vi} and \ion{Ca}{vi}, have two distinct levels
with multiple decay routes, while the remaining three ions have a
single level with multiple decay routes, thus in total there are nine distinct
consistency checks on the combined STIS--UVES wavelength scale. For
each of the nine cases, the decay routes are a direct decay to the
ground term and an indirect route via an intermediate
level.\footnote{For \ion{Ne}{iii} the direct route is 
  to the $^3P_1$ level of the ground term while the indirect route is
  to the $^3P_2$ level. The $^3P_1$--$^3P_2$ transition has been
  accurately measured by \citet{feucht97} and so this value was used
  to complete the calculation for \ion{Ne}{iii}.} The
checks consisted of measuring the wavelengths of the transitions of
the indirect route and using these to predict a wavelength for the
direct route. Expressing the difference (predicted $-$ observed) in wavelengths as a velocity
we find the average value to be $+2.6$~\kms\ with a standard
deviation of $3.6$~\kms, thus there is no systematic difference
between short and long wavelength measurements. 

\begin{figure}[h]
\plotone{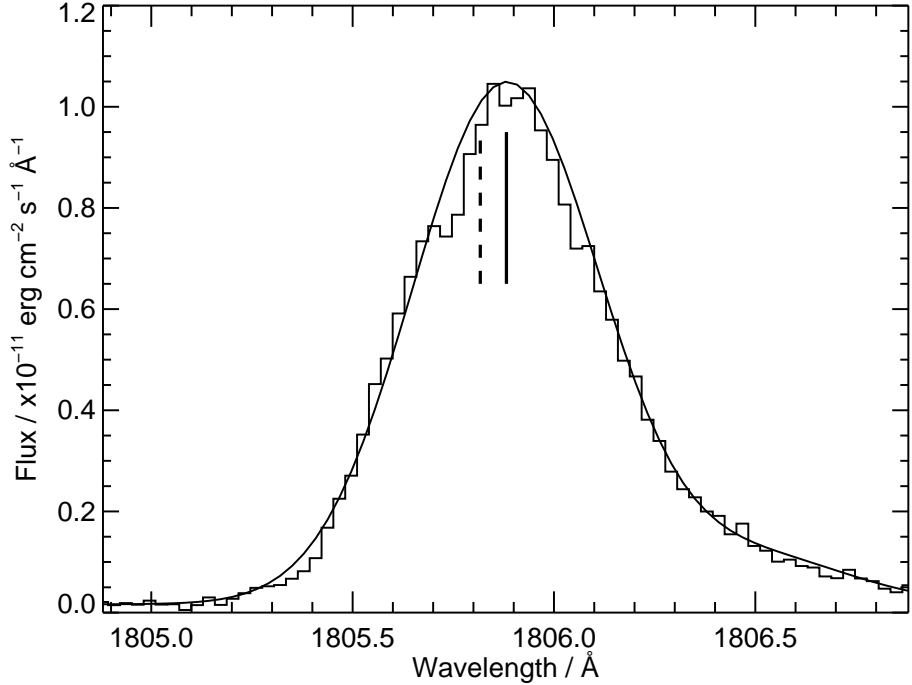}
\caption{Line profile of the \ion{Mg}{vi} \lam1805.9+\lam1806.3 self-blend from the STIS
  spectrum with a two Gaussian fit over-plotted. The solid vertical line indicates the centroid of the 
  fitted Gaussian for \lam1805.9 and the dashed vertical line indicates the centroid
  position required to satisfy the consistency check discussed in
  Sect.~\ref{sect.consistency}.}
\label{fig.mg6}
\end{figure}

The worst agreement between observed and predicted wavelengths is for
\ion{Mg}{vi}, and it is significantly outside of the combined 1$\sigma$
errors of other sources of error discussed in Sects.~\ref{sect.stis}
and \ref{sect.uves}.  Agreement between the observed and predicted
wavelengths is an absolute requirement necessary for the integrity of the
present work. For this reason we introduce an additional error,
expressed as a velocity, that forces the \ion{Mg}{vi} wavelengths to
agree with each other within the 1$\sigma$
errors of the wavelengths. This velocity error is found to be 5.5~\kms,
and is added in quadrature to the other sources of error discussed
earlier. To illustrate the magnitude of the velocity error we show in
Fig.~\ref{fig.mg6} the \ion{Mg}{vi} \lam1805.9 line profile with the
measured line centroid indicated together with the centroid position
derived from combining the wavelengths of the two lines from the
alternative decay route.

We believe that this additional error source is not due to the instruments, but instead is due
to the velocity and density structure of the nebula and the different
sensitivities of emission lines to density. For example, if an ion's emission lines
arise from two plasma components at $\pm 10$~\kms\ and these
components have different densities, then a pair of emission lines that
are density sensitive will display different line profiles that may
lead to different centroid positions for the lines.
Since the resolution and
sensitivity, particularly of the STIS spectra, are not high enough to
clearly resolve detailed structure in the line profiles then this
velocity structure may be smoothed over, and only revealed through
anomalous centroids for the lines.

\section{Wavelengths and energy levels}

The following sections give details on all the forbidden lines
measured for the present work. Also included for some ions are intercombination
lines. Generally in laboratory or astronomy literature wavelengths
above 2000~\AA\ are given as air wavelengths  whereas those below
2000~\AA\ are given as vacuum wavelengths. For the present work we
will only use vacuum wavelengths unless otherwise stated. For
conversions between air and vacuum wavelengths we use
the IDL routines VACTOAIR and AIRTOVAC that are distributed through
the Astronomy IDL library and use the formula given by \citet{morton91}.

Table~\ref{tbl.wavelengths} presents the wavelengths measured in the
present work given in both vacuum and air forms. 1$\sigma$ errors on
the measurements, calculated as described in Sect.~\ref{sect.abs.wvl}
are also given. The 
NIST database gives energy values for all ions, and the wavelengths
derived from these values using version 3 of the online database \citep{ralchenko08} are
presented. Some previous wavelength measurements from astrophysical sources
are also presented. \citet{thackeray74} and \citet{thackeray77} give
optical wavelengths derived from RR Tel; \citet{penston83} give
ultraviolet wavelengths measured from IUE spectra of RR Tel; and
\citet{jordan98} presented wavelengths for some lines from HST/GHRS
RR Tel spectra. Many of the intercombination lines and some of the
higher ionization forbidden lines have been measured in solar spectra,
and we include values from \citet{doschek76}, \citet{doschek77},
\citet{sandlin77} and \citet{curdt04}. Each of these spectral atlases
was obtained above the solar limb where the plasma can be reasonably
considered to be at rest. Emission lines observed on the solar disk
are well known to show systematic 
velocity shifts \citep{doschek76a,peter99} and so a spectral atlas such as that of
\citet{curdt01} is not useful for determining rest wavelengths.

In some parts of the text we refer to the atomic models from the
CHIANTI database. CHIANTI gives atomic data for modeling the emission
processes of forbidden lines which can be useful for determining the
detectability of a line if another line from the ion is known.

\subsection{Beryllium isoelectronic sequence}

For the present work we consider only the intercombination line $2s^2$
$^1S_0$ -- $2s2p$ $^3P_1$ and the forbidden line $2s^2$
$^1S_0$ -- $2s2p$ $^3P_2$. Other lines from the Be-like ions are found
in the spectra but these are either resonance lines or recombination
lines. 

\subsubsection{N\,IV}\label{sect.n4}

The intercombination line, \lam1486.5, is strong and lies at the edge of two
spectral orders in the SW spectrum. The line has an asymmetric line
profile with an extended long wavelength wing and has been fit with
two Gaussians forced to have the same width. The stronger, short
wavelength component is assumed to be at the rest wavelength of the
system and the average wavelength from the two spectral orders was
used to derive the wavelength given in
Table~\ref{tbl.wavelengths}. The wavelength is in good agreement with
the NIST wavelength, the previous RR Tel measurement of
\citet{penston83}, and the solar measurements of \citet{doschek76} and \citet{sandlin77}.

The nearby forbidden line, \lam1483.3, is measured but shows a
significant redshift relative to the NIST wavelength: the measured
wavelength is $1483.385\pm 0.031$~\AA\ compared to the NIST value of
1483.321~\AA, equating to a velocity shift of 13~\kms. This is
consistent with the corresponding forbidden line of \ion{C}{iii}
(Sect.~\ref{sect.c3}) and
suggests that it too arises from the redshifted plasma component.

\subsubsection{O\,V}

The   forbidden and intercombination lines lie either side of the very broad
interstellar absorption line of \ion{H}{i} \lya\ at 1213.8 and
1218.3~\AA, respectively.  The forbidden line was not reported by
\citet{penston83} from IUE spectra of RR Tel, but is clearly seen in
the STIS spectrum. It lies on the sloping wing of the \lya\ absorption
feature and thus the long wavelength wing will be more absorbed than
the short wavelength wing which may create a false blueshift for the
line profile. Due to the higher ionization potential of \ion{O}{v} we
do not expect the forbidden line to be emitted from the low density
plasma component as was found for \ion{C}{iii} and \ion{N}{iv}. The
measured centroid lies blueward of the only 
previous measurement of the \lam1213.8 line \citep{sandlin77}, but the
wavelength is in excellent agreement with the NIST wavelength.

As with other intercombination lines, \lam1218.3 shows an asymmetric
profile with an extended long wavelength wing. The effect is more
pronounced in this case, however because of the \lya\ absorption on
the short wavelength side of the profile. The line has been fit with
two Gaussians forced to have equal width and with the short wavelength
side of the profile partly masked off to prevent the \lya\ absorption
distorting the fit. The strongest of the fitted Gaussians is taken to
correspond to the rest component of the system and yields the
wavelength given in Table~\ref{tbl.wavelengths} which is in good
agreement with the previous measurements of \citet{doschek76}, \citet{sandlin77} and
\citet{penston83}.

\subsection{Boron isoelectronic sequence}

The intercombination transitions, $2s^22p$ $^2P_J$ -- $2s2p^2$
$^4P_{J^\prime}$, of the boron sequence are found in the STIS spectrum
for \ion{C}{ii}, \ion{N}{iii} and \ion{O}{iv}. The \ion{F}{v} lines
are expected around 1167~\AA\ but can not be identified. 

\subsubsection{C\,II}

All five of the \ion{C}{ii} intercombination lines are found in the RR
Tel spectrum. The four strongest lines all show asymmetric profiles with
enhanced long wavelength wings, similar to other intercombination
lines in the spectrum, and they were fit with two Gaussians forced to
have the same width. The wavelength of the stronger, short wavelength
component was used to generate the rest wavelengths in Table~\ref{tbl.wavelengths}. For the weakest transition, $^2P_{1/2}$ -- 
$^4P_{1/2}$, the asymmetry is not pronounced due to the lower
signal-to-noise and a single Gaussian fit was used. The 3/2--3/2
transition at 2327.6~\AA\ has a much more extended long wavelength
wing than the other lines which is likely due to the \ion{Fe}{ii}
a~$^6D_{5/2}$--z~$^6P_{3/2}$ transition at 2328.111~\AA. This part of
the profile was thus not included in the two Gaussian fit to the
\ion{C}{ii} line. 

The \ion{C}{ii} intercombination lines have been previously measured in off-limb solar
spectra by \citet{doschek77}, and in the IUE spectrum of RR Tel by
\citet{penston83} and the comparison in Table~\ref{tbl.wavelengths}
shows good agreement except for the \lam2324.3 and \lam2328.9 lines,
where the \citet{penston83} wavelengths are significantly shorter than
the current wavelengths. (They are also significantly shorter than the
Doschek \& Feldman~1977 wavelengths.)

One discrepancy that is present for all three sets of measurements
comes from deriving the ground $^2P_J$ level splitting by using the
two pairs of transitions $^2P_{1/2,3/2}$--$^4P_{1/2}$ and
$^2P_{1/2,3/2}$--$^4P_{3/2}$. For the current wavelength
measurements\footnote{Since the \ion{C}{ii} lines all lie within a
  single echelle order of the spectrum, and the lines are used to
  derive level splittings (thus the absolute wavelength is not
  needed), then the only error component for the 
  wavelengths that is
  relevant is the Gaussian fitting error which is much smaller than
  the error given in Table~\ref{tbl.wavelengths}.}
the latter transition pair yields a splitting of $62.79\pm
0.16$~cm$^{-1}$, while the former yields $63.89\pm 0.10$~cm$^{-1}$,
neither of which is consistent with the very accurate value of
63.39~cm$^{-1}$ from \citet{cooksy86}. This may reflect short scale
inhomogeneities in the STIS wavelength or suggest that the
wavelengths are affected by the non-Gaussian shapes of the line
profiles. 

\subsubsection{N\,III}

The five intercombination lines are found within 8~\AA\ of each other
and all have very similar line widths of around 30~\kms, suggesting
they are unblended. The two strongest transitions, \lam1749.7 and
\lam1752.1, clearly show asymmetric profiles similar to other
intercombination lines in the spectrum and have been fit with two
Gaussians forced to have the same width. The stronger, shorter
wavelength Gaussians are used to derive the rest wavelengths in
Table~\ref{tbl.wavelengths}. Single Gaussian fits were used for the
remaining transitions.
The measured wavelengths are in excellent
agreement with the solar measurements of \citet{doschek76} and 
\citet{sandlin77}. The agreement with the wavelengths of
\citet{penston83} from IUE spectra of RR Tel are slightly less good,
but consistent within the uncertainties.

\subsubsection{O\,IV}\label{sect.o4}

The \ion{O}{iv} intercombination lines are very strong in RR Tel and
have been studied in some detail by previous authors
\citep{harper99,keenan02}. As with other strong intercombination lines
such as \ion{C}{iii} \lam1908.7, \ion{O}{iii} \lam\lam1660.8, 1666.2
and \ion{Si}{iii} \lam1892.0 the \ion{O}{iv} lines have asymmetric
profiles with the long wavelength wings of the lines being more
extended than the short wavelength wings. They have been fit with two
Gaussians forced to have the same width, with the stronger component
being taken as that of the rest component of the nebula. 

Three previous measurements of the lines' wavelengths have been made
from RR Tel spectra by \citet{penston83}, \citet{harper99} and
\citet{keenan02} from IUE, HST/GHRS and HST/STIS spectra,
respectively, the latter
work using the same spectra as used here. In addition the lines have
also been measured from solar spectra by \citet{doschek76} and \citet{sandlin77}. Comparing
the results in Table~\ref{tbl.wavelengths} it is clear that the
\citet{keenan02} wavelengths are blueshifted relative to the other
results by around 0.02--0.06~\AA. We believe this is because the
authors used the nearby \ion{Si}{iv} resonance lines to determine a
rest wavelength scale. We find that these lines, like other resonance
lines in the spectrum, are redshifted relative to the system's radial
velocity by 10~\kms\ and so lead to the wavelength offset for the
\ion{O}{iv} lines.

The \lam1404.8 line is known to blend with a \ion{S}{iv} transition,
but \citet{harper99} demonstrated that this line contributes only 4\%\
to the measured feature's flux and so the measured line centroid is a
reliable measure of the \ion{O}{iv} line's wavelength.

The present wavelengths tend to be midway between those of
\citet{penston83} and \citet{harper99}, the latter's results being
close to those of \citet{sandlin77}. The wavelength separations of the
lines are consistent between all of the measurements to within a few
m\AA. The NIST wavelengths show significant discrepancies with the
astrophysical wavelengths and should be revised.

\subsection{Carbon isoelectronic sequence}

There are six forbidden lines for carbon-like ions, and the strongest
are $^3P_{1,2}$--$^1D_2$, $^1D_2$--$^1S_0$ and $^3P_1$--$^1S_0$. No
lines can be identified from \ion{F}{iv}  but otherwise all the ions
from \ion{N}{ii} to \ion{Mg}{vii} are represented. The \ion{Mg}{vii}
forbidden lines are found to be much stronger in the symbiotic star AG
Draconis and new rest wavelengths for the \ion{Mg}{vii} lines were
discussed in \citet{young06}.

\subsubsection{N\,II}

The model presented in Appendix~\ref{app.model}
suggests that \ion{N}{ii} \lam6585.3, which is emitted from the $^1D_2$ level,
principally comes from the low density plasma component, whereas
\lam5756.2, which is emitted from the higher energy $^1S_0$ level,
principally comes from the high density plasma component.
The \lam6585.3 line profile shows
three distinct components, the strongest of which corresponds to the redshifted
component seen in \ion{O}{ii} \lam3729.9, and so is consistent with the
model. The additional components are at the rest
wavelength of the system and  at a velocity of around
$-30$~\kms\ (this is a broad component). The profile thus seems to be
midway between \ion{O}{ii} \lam3729.9 and \ion{O}{iii} \lam5008.2. 
The weaker $^3P_{2}$--$^1D_2$ transition at 6549.9~\AA\
shows a similar structure (as expected since the lines share the same
upper level) but the line seems to be affected by a blend in the short
wavelength wing. We do not include these two transitions in
Table~\ref{tbl.wavelengths} on account of the complexity of the line profiles.

As expected from the emission model of Appendix~\ref{app.model}, the 
 $^1D_2$--$^1S_0$ emission line at 5756.2~\AA\ has a much simpler
 profile than \lam6585.3. Gone is the broad, blueshifted component at $-30$~\kms,
while the redshifted component is weaker than the rest
component. 
Fitting the line with two Gaussians yields the wavelength
for the rest component in Table~\ref{tbl.wavelengths}, which is in
good agreement with the NIST wavelength and the value of
\citet{bowen60}.

The $^3P_1$--$^1S_0$ transition at 3063.7~\AA\ is partly blended with a
\ion{O}{iv} recombination line at 3064.3~\AA\ but a weak component in
the short wavelength wing of this line can be identified as the \ion{N}{ii}
transition. A two Gaussian fit was performed and the wavelength of the
weak component is given in Table~\ref{tbl.wavelengths}. Since
\lam3063.7 is emitted from the same upper level as \lam5756.2 then it
will be expected to show the same two component structure as this
line. It is not possible to resolve the redshifted component on
account of the blending \ion{O}{iv} line. For this reason the
wavelength given in Table~\ref{tbl.wavelengths} should be treated with
caution and the more accurate value of \citet{bowen60} is preferred.

The intercombination lines occur at 2139.7 and 2143.5~\AA\ for
\ion{N}{ii}, but the latter is blended with a \ion{Fe}{vii} transition
\citep[see the discussion in][]{young05a}
and the wavelength can not be accurately estimated.  \lam2139.7 shows
an extended long wavelength wing like other intercombination lines in
the RR Tel spectrum and the feature has been fit with two Gaussians
forced to have the same width. The wavelength of the stronger, short
wavelength component is given in Table~\ref{tbl.wavelengths} and is
found to be in excellent agreement with the NIST wavelength but
discrepant with the measurement of \citet{penston83} from IUE spectra
of RR Tel.

\subsubsection{Ne\,V}

Six \ion{Ne}{v} forbidden lines and one intercombination line are found in the STIS and UVES
spectra. The longest wavelength lines are the decays of the $^1D_2$
level to the $^3P_1$ and $^3P_2$ levels, 
giving two strong lines at 3347.0 and 3427.0~\AA, respectively. The
UVES wavelengths agree with those of \citet{bowen60} within the
uncertainties, while the separation of the two UVES lines implies a
$^3P_1$--$^3P_2$ separation of $698.3\pm 0.8$~cm$^{-1}$ in good agreement  with the measurement of $698.242 \pm 0.010$ of
\citet{feucht97} from infrared spectra. The $^1D_2$ level also yields
a weak decay to $^3P_0$ at 3301.3~\AA\ but this is blended with a much stronger
\ion{O}{iii} line at 3300.3~\AA\ and can not be measured.

The $^1S_0$ level also decays to the $^3P_1$ and $^3P_2$ levels, giving two lines at
1574.7 and 1592.2~\AA. The latter is a weak line and was not measured
in IUE spectra of RR Tel \citep{penston83}. The separation of the two
lines implies a $^3P_1$--$^3P_2$ separation of $698.6\pm 2.2$ which
is in good agreement with the measurement of $698.242 \pm 0.010$ of
\citet{feucht97}. The \lam1574.7 wavelength is
in very good agreement with that of \citet{penston83}.

The $^1D_2$--$^1S_0$ transition is found in the STIS spectra at
2974.0~\AA\ and agreement within the uncertainties is found with the
previous measurement of
\citet{penston83}. Combining the measured wavelength of this line
with that of \lam3346.8, yields a predicted wavelength of
1574.699~\AA\ for the $^3P_1$ -- $^1S_0$ which is 5.3~\kms\ longward
of the measured wavelength for this transition (see Sect.~\ref{sect.consistency}).

The intercombination transition, $2s^22p^2$ $^3P_2$ -- $2s2p^3$
$^5S_2$, occurs at 1145.6~\AA\ and is the shortest wavelength line
found in the STIS spectrum. The measured wavelength is a little
shorter than the solar measurements presented by \citet{sandlin77} and
\citet{young05b}, but consistent within the uncertainties.

\subsubsection{Na\,VI}

Measurements of the \ion{Na}{vi} forbidden lines have not previously
been reported in the literature, and the estimated wavelengths of
\citet{bowen60} have large uncertainties. However, \citet{edlen72} provided
 calculated energy levels that yield accurate
wavelengths. Note that these energies are found to be significantly more accurate
than those contained in the NIST database. Four \ion{Na}{vi} forbidden
lines are found in the RR Tel
STIS spectra, and the two strongest  are the $^3P_{1,2}$--$^1D_2$ transitions at
2872.65 and 2971.79~\AA, respectively, which are close to the
predicted wavelengths of \citet{edlen72}: 2872.59 and 2971.65~\AA,
respectively. The  separation of the lines implies a
$^3P_1$--$^3P_2$ splitting of $1161.3\pm 1.0$~cm$^{-1}$, in good
agreement with the infrared measurement of \citet{feucht97} who found
$1161.36\pm 0.12$~cm$^{-1}$.

The $^1D_2$--$^1S_0$ and $^3P_1$--$^1S_0$ transitions are both weak
and \citet{edlen72} predicts wavelengths of 2569.71 and 1356.36~\AA,
respectively. The former is close to a broad, weak line in the STIS
spectrum at 2569.59~\AA\ that we identify with the \ion{Na}{vi}
transition. A clump of five emission lines is found in the STIS
spectrum at 1356~\AA, and one at 1356.32~\AA\ is close to the
\citet{edlen72} wavelength and has a line width consistent with the
other \ion{Na}{vi} lines. Another line at 1355.94~\AA\ has a similar
flux and width, and thus is another potential candidate for the
\ion{Na}{vi} transition. However, by using the measured wavelengths of
the $^3P_{1}$--$^1D_2$ and $^1D_2$--$^1S_0$ transitions we can predict a
wavelength of $1356.34\pm 0.03$~\AA\ for the $^3P_1$--$^1S_0$ which is
consistent with the observed 1356.32~\AA\ line.

\subsection{Nitrogen isoelectronic sequence}

There are eight forbidden lines from nitrogen-like ions and all
except the weak $^2D_{5/2}$--$^2P_{1/2}$ transition are potentially measurable
in RR Tel. No lines can be found from \ion{F}{iii}, but all other ions
are represented up to \ion{Mg}{vi}. 

\subsubsection{O\,II}

The $^2D$--$^2P$ transitions lie between 7321 and 7333~\AA\ and so
outside the UVES wavelength range, while the $^4S$--$^2D$ transitions
were discussed in Sect.~\ref{sect.cpts} where they were found to
be emitted from a redshifted plasma component. 

The $^4S_{3/2}$--$^2P_{1/2,3/2}$ lines occur at 2471.0 and 2471.1~\AA\
and are found blended in a single spectral feature in the STIS
spectrum. The CHIANTI atomic model predicts that \lam2471.1 should be
around four times stronger than the companion line, and also that both
lines are significantly more sensitive to high densities than the
\lam\lam3727.1,3729.9 line pair. The latter point means that the
$^4S_{3/2}$--$^2P_{1/2,3/2}$ transitions may have a significant
component from the rest component of the plasma (see also
Appendix~\ref{app.model}), unlike the $^4S$--$^2D$ transitions.  The observed line profile
does not show any clear asymmetry nor any evidence of extended
wings. Fitting it with a single Gaussian and assuming it is entirely
due to the $^4S_{3/2}$--$^2P_{3/2}$ transition yields  the wavelength
shown in Table~\ref{tbl.wavelengths}, which is discrepant with both
the NIST wavelength and the \citet{bowen60} wavelength. Given the
uncertainty over the contribution from the redshifted, low density
plasma and degree of blending we advise the reader to treat the
present \lam2471 wavelength measurement with caution.

\subsubsection{Ne\,IV}\label{sect.ne4}

Wavelengths and energy levels of \ion{Ne}{iv} were assessed by
\citet{kramida99}. The $^2D$--$^2P$ forbidden transitions lie between
4715 and 4727~\AA\ and so are not found in the present UVES spectra,
thus the wavelengths of \citet{bowen60} still represent the best
measurements of these lines. The $^4S_{3/2}$--$^2P_{1/2,3/2}$
transitions at 1601.5 and 1601.7~\AA\ are
close in wavelength and have not been resolved in the previous
measurements of \citet{sandlin77} and \citet{penston83}. The line in
the STIS spectrum is clearly asymmetric, suggesting two lines with the
weaker lying in the long wavelength wing of the stronger line. Fitting
the feature with two Gaussians forced to have the same width yields
the wavelengths listed in Table~\ref{tbl.wavelengths}. 

The
$^4S_{3/2}$--$^2D_{3/2,5/2}$ transitions are separated by around
3~\AA, but both are partly blended with other, narrow lines. Simultaneous two
Gaussian fits were performed to each feature to resolve the
components. The stronger of the two \ion{Ne}{iv} lines, \lam2422.4,
has an unidentified narrow line in the short wavelength wing, which is
likely a \ion{Fe}{ii} transition. The weaker \ion{Ne}{iv} line,
\lam2425.0, also has a narrow line in the short wavelength wing that
can be identified with the \ion{Fe}{ii} $b~^4F_{9/2}$ -- $y~^4G_{11/2}$ transition
(\lam2424.883). The widths of the two \ion{Ne}{iv} lines are 43 and
41~\kms, respectively, which are in very good agreement with the width
of 41~\kms\ found for the $^4S_{3/2}$--$^2P_{1/2,3/2}$ transitions,
giving confidence in the two Gaussian fit employed for these
lines. 

The analysis presented in Appendix~\ref{app.model} suggests that the
\ion{Ne}{iv} $^4S_{3/2}$--$^2D_{3/2,5/2}$  transitions may be principally formed
in the redshifted, low density plasma component of the nebula, unlike
the $^4S_{3/2}$--$^2P_{1/2,3/2}$ transitions. This may explain the
wavelength differences compared to \citet{penston83} and
\citet{jordan98}, although this would imply the low density plasma
component was not present at the time of these earlier RR Tel
observations.

Since the four decays to the ground level described above directly
yield the energies of the four excited levels in the ground
configuration, one can then derive wavelengths for the four $^2D$--$^2P$
transitions and compare with the wavelengths presented by
\citet{bowen60}. The uncertainties on the derived wavelengths are
around 0.35~\AA, significantly larger than those of \citet{bowen60}
which are 0.04~\AA. We find agreement within these uncertainties for
all of the transitions except $^2D_{3/2}$--$^2P_{3/2}$ for which the
derived air wavelength is 4723.75~\AA\ and the \citet{bowen60}
wavelength is 4724.15~\AA. Note that this discrepancy could be
explained if the $^4S_{3/2}$--$^2D_{3/2,5/2}$ transitions are formed
in the redshifted, low density plasma component.

\subsubsection{Na\,V}\label{sect.na5}

The four $^2D$--$^2P$ transitions are expected to lie between 4012 and
4026~\AA\ and so are not found in the UVES spectra. The
$^4S_{3/2}$--$^2D_{3/2,5/2}$ and $^4S_{3/2}$--$^2P_{1/2,3/2}$
transitions occur
at ultraviolet wavelengths and three of the transitions can be identified
in the STIS spectra. \cite{edlen72} provided calculated energies for
the ground levels of \ion{Na}{v} and these yield predicted wavelengths
for the  $^4S_{3/2}$--$^2D_{3/2,5/2}$ transitions of 2067.85 and
2069.81~\AA. A line is found at the former wavelength but it is
blended with the $n=33$ member of the \ion{He}{ii} Fowler
series. However, by comparison with other members of the Fowler series
it is clear that \ion{Na}{v} provides the dominant contribution to the
blend and so we associate the measured wavelength with
\ion{Na}{v}. The $^4S_{3/2}$--$^2D_{3/2}$ \ion{Na}{v} transition is
also blended with a Fowler series line, in this case the $n=31$
member, and it lies in the short wavelength wing of a much
stronger line that we believe is due to \ion{O}{vi}. The strength of the
\ion{Na}{v}--\ion{He}{ii} blend is consistent with the line
predominantly arising from \ion{He}{ii} and so we do not associate the
measured wavelength with \ion{Na}{v}.

As for \ion{Ne}{iv}, Appendix~\ref{app.model} suggests that the
$^4S$--$^2D$ transitions may be predominantly formed in the
redshifted, low density plasma component, therefore readers are
recommended to treat the rest wavelength for \lam2069.9 in
Table~\ref{tbl.wavelengths} with caution.

\citet{penston83}
identified a line at 1365.37~\AA\ that the authors identified with
both of the $^4S_{3/2}$--$^2P_{1/2,3/2}$ transitions. The STIS spectra
clearly resolve both components at 1365.39 and 1366.08~\AA, with the
former being stronger by a factor three thus the \citet{penston83}
wavelength seems to correspond only to the $^4S_{3/2}$--$^2P_{3/2}$
transition. The calculated energy values of \citet{edlen72} yield
predictions of 1365.44 and 1366.08~\AA\ for the two transitions, in
good agreement with the STIS measurements.

\subsubsection{Mg VI}

The four $^2D_{3/2,5/2}$--$^2P_{3/2,1/2}$ transitions are found in the
UVES spectra between 3488 and 3504~\AA. The 5/2--1/2 transition is
weak and difficult to measure so is not listed in
Table~\ref{tbl.wavelengths}. Wavelengths were given for all four lines
by \citet{bowen60} but these were obtained by calculation and are only
accurate to $\pm$~3~\AA. The three lines observed in the UVES spectra
allow the splittings of the $^2P$ and $^2D$ terms to be derived and we
obtain values of $14.7\pm 0.8$ and $108.7\pm 0.8$~cm$^{-1}$ for the
$^2D$ and $^2P$ terms, respectively.

The $^4S_{3/2}$--$^2P_{3/2,1/2}$ transitions give rise to two strong
lines at 1190.0 and 1191.6~\AA, respectively, with widths of 69 and
75~\kms. \lam1190.0 is partly blended with \ion{Mg}{vii} 
\lam1189.9, but based on the flux of the unblended \ion{Mg}{vii} \lam2629
line we estimate this contributes less than 1\%. The separation of the
\lam1190.0 and 1191.6 lines yields a separation for the $^2P$ levels of
$109.2 \pm 2.4$~cm$^{-1}$, in good agreement with the previous
determination. 

The $^4S_{3/2}$--$^2D_{3/2,5/2}$ lines are close in wavelength and the
observed feature in RR Tel shows a strong line with an extended long
wavelength wing (Fig.~\ref{fig.mg6}). A two Gaussian fit was performed, forcing the two
lines to have the same width, however the resulting wavelengths are not
consistent with the separation of the $^2D_{3/2,5/2}$ levels obtained
above: the implied energy separation is $19.2\pm 0.6$~cm$^{-1}$. The
\lam1190.0 and \lam1191.6 lines both show some structure in their line profiles
beyond a simple Gaussian shape and this is also seen in \lam1806, it
thus seems that the weak $^4S_{3/2}$--$^2D_{5/2}$ transition (which is
expected to be around a factor ten weaker than its neighbor) is not
correctly extracted by assuming a two Gaussian fit. In
Table~\ref{tbl.wavelengths} we list only the $^4S_{3/2}$--$^2D_{3/2}$
transition.

As discussed in Sect.~\ref{sect.consistency}, the \ion{Mg}{vi} lines
are useful for checking the wavelength scales of the UVES and STIS
spectra, however a significant discrepancy was found that led to an
additional source of  uncertainty in the present analysis.
The measured wavelengths of \lam\lam3503.2, 3489.9 and 1805.9 yield
predictions for the short wavelength lines of 1190.068 and
1191.610~\AA\ that are up to 0.028~\AA\ different from the
measurements -- a difference of 7.1~\kms.

\subsection{Oxygen isoelectronic sequence}

\subsubsection{Ne III}

The strongest line from \ion{Ne}{iii} is the $^3P_2$--$^1D_2$
transition at 3869.8~\AA\ and we find that it has a
distinctive `shoulder' on the long wavelength side of the profile
that is around half the strength of the main line. We believe this is
emission from the red-shifted plasma component that is prominent in the
cooler ions (Sect.~\ref{sect.cpts} and Appendix~\ref{app.model}). The line has thus been fit with two Gaussians and the
shorter wavelength Gaussian gives the wavelength quoted in
Table~\ref{tbl.wavelengths}, which is in good agreement with the value
of \citet{bowen60}.

The $^1D_2$--$^1S_0$
transition, \lam3343.4, is much weaker than \lam3869.8 and shows an extended long
wavelength wing that is relatively less intense than the shoulder of
the \lam3869.8 line. \lam3343.4 is more sensitive to high densities than
\lam3869.8 and so the weaker wing found for this line is due to the
lower density of the red-shifted plasma component.

The $^3P_1$--$^1S_0$ transition is found at 1814.65~\AA\ and there is
no evidence of a long wavelength wing to the profile, although the
signal in the line is  weaker than the longer wavelength
lines.

The three \ion{Ne}{iii} lines allow a consistency check to be
performed on the UVES and STIS wavelength scales
(Sect.~\ref{sect.consistency}), although it is necessary to use the
$^3P_2$--$^3P_1$ separation of \citet{feucht97} to complete the
calculation. We find a predicted wavelength for the $^3P_1$--$^1S_0$
transition of 1814.636, within 0.01~\AA\ of the STIS measurement.

\subsubsection{Na\,IV}

The \ion{Na}{iv} $^3P_{1,2}$--$^1D_2$ transitions are found in the
UVES spectra at 3242.7 and 3363.3~\AA\ and the measured wavelengths
are consistent with those measured by \citet{bowen60}, although the
UVES spectra have smaller uncertainties. 
Transitions
from the $^1S_0$ level are expected in the STIS wavelength range, but
none can be identified.

\subsubsection{Mg\,V}

The \ion{Mg}{v} forbidden lines all lie in the ultraviolet part of the
spectrum and so \citet{bowen60} was only able to give approximate
wavelengths based on calculations and extrapolation along the
isoelectronic sequence. Four lines are found in the STIS spectrum, one
of which has not previously been reported.

\citet{penston83} reported the $^3P_{1,2}$--$^1D_2$ transitions from
the IUE spectrum of RR Tel and the STIS wavelengths are in excellent
agreement (Table~\ref{tbl.wavelengths}). The separation of the two
lines implies a splitting of the $^3P_{1,2}$ levels of $1782.7\pm
1.0$~cm$^{-1}$, in good agreement with the measured infrared value of
$1782.58\pm 0.20$~cm$^{-1}$ of \citet{feucht97}.
The infrared measurement of the $^3P_1$--$^3P_0$ splitting
\citep[also][]{feucht97} can be
used to predict a wavelength for the $^3P_{0}$--$^1D_2$ transition of
2993.84~\AA. This is close to a line at 2993.73~\AA\ that has a width
consistent with the other members of the multiplet, but the wavelength
discrepancy is outside the uncertainties and the strength of the
observed line is also larger than expected so we do not make the
identification.

The $^1D_2$--$^1S_0$ transition at 2417.6~\AA\ has not previously been
reported in the literature, but is clearly seen in the STIS
spectrum and the wavelength is given in Table~\ref{tbl.wavelengths}.
The $^3P_1$--$^1D_2$ transition at 1324.4~\AA\ has been 
measured in solar spectra by \citet{sandlin77} and in IUE spectra of
RR Tel by \citet{penston83} and both measurements are in good
agreement with the STIS wavelength (Table~\ref{tbl.wavelengths}).

Combining the $^1D_2$--$^1S_0$ and $^3P_1$--$^1D_2$ wavelengths yields
a predicted wavelength of 1324.427~\AA\ which is within 2~\kms\ of the
measured position, and confirms the identification of the \lam2417.6 line.

Finally we note that there are significant  discrepancies  between the
wavelengths derived from the NIST energy levels and the present
measurements, suggesting the NIST database needs to be updated.

\subsubsection{Al\,VI}

Two \ion{Al}{vi} lines are found in the STIS spectrum: the
$^3P_{2,1}$--$^1D_2$ transitions at 2430.2 and 2603.1~\AA,
respectively. The former was previously identified by \citet{jordan98}
and the present wavelength is in good agreement with their
value. \lam2603.1 has not previously been reported.

\subsection{Magnesium isoelectronic sequence}

\ion{Al}{ii} and \ion{Si}{iii} lines were used as wavelength fiducials
(Sects.~\ref{sect.al2} and \ref{sect.si3}). The intercombination
transition, $3s^2$ $^1S_0$ -- $3s3p$ $^3P_1$, 
is found for both \ion{P}{iv} and \ion{S}{v} and the lines are
discussed below.

\subsubsection{P\,IV}

The \ion{P}{iv} intercombination line was first measured in the
laboratory by \citet{robinson37} and later by \citet{zetterberg77},
and their values of 1467.424 and 1467.427~\AA, respectively, are in good
agreement with the present measurement (Table~\ref{tbl.wavelengths}). The line has also been
measured from solar spectra by \citet{sandlin77} and from IUE spectra
of RR Tel by \citet{penston83}, and agreement is again good.

\subsubsection{S\,V}

The \ion{S}{v} intercombination line is close to a strong interstellar
absorption line of \ion{N}{i} \lam1199.55, but the line profile is not
affected and a good measurement of the line centroid can be made. The
wavelength agrees with the solar measurement of \citet{sandlin77} and
the \citet{penston83} value from the IUE spectra of RR Tel.

\subsection{Aluminium isoelectronic sequence}

The intercombination transitions $3s^23p$ $^2P_J$ -- $3s3p^2$
$^4P_{J^\prime}$ are the only ones expected to appear in the RR Tel
spectrum, and the \ion{Si}{ii} lines between 2329 and 2351~\AA\ were
used as wavelength fiducials (Sect.~\ref{sect.si2}). Lines from
\ion{S}{iv} are clearly seen in the spectrum and discussed below, but
no lines from \ion{P}{iii} or \ion{Cl}{v} can be found. The
\ion{Ar}{vi} lines lie below the short wavelength limit of the STIS
spectrum.

\subsubsection{S\,IV}

The \ion{S}{iv} transitions lie close in wavelength to the stronger 
intercombination transitions of \ion{O}{iv} (Sect.~\ref{sect.o4}), and
the STIS RR Tel lines have previously been studied by
\citet{keenan02}. The 1/2--1/2 transition at 1404.8~\AA\ is blended with one of the
\ion{O}{iv} transitions and \citet{keenan02} found that the \ion{S}{iv} transition
contributes less than 2\%\ to the observed line's intensity so   the
line can not be used to determine a rest wavelength.  

The strongest transitions, \lam1406.0 and \lam1416.9 are both unblended
and well-observed in the spectrum. As with other intercombination
lines in the RR Tel spectrum, they show asymmetric line profiles and
have been fitted with two Gaussians forced to have the same width. 
The
stronger, shorter wavelength component is used to derive the rest
wavelengths, which are in good agreement with previous measurements
(Table~\ref{tbl.wavelengths}) 
except for \citet{keenan02}. As mentioned in Sect.~\ref{sect.o4}, this
is probably due to the use of the \ion{Si}{iv} resonance lines as
wavelength fiducials by these authors.

The 1/2--3/2 transition at 1398.0~\AA\ is extremely weak
but can be measured, although only a single Gaussian fit was used due to the
low signal-to-noise. The \lam1398.0 line has only been measured
previously from  RR Tel spectra, and there is a significant discrepancy between
the present wavelength and \citet{penston83} value. It is possible
that the \citet{penston83} measurement was affected by the nearby
\ion{Fe}{ii} \lam1397.845 line which is clearly resolved in the STIS
spectrum, but may have blended with the \ion{S}{iv} line in the IUE spectrum.

The 3/2--1/2 transition at 1423.8~\AA\ is blended with a broad
spectral feature, and there is also a narrow \ion{Fe}{ii} line
nearby. Performing a three Gaussian fit yields the wavelength for the
\ion{S}{iv} line given in Table~\ref{tbl.wavelengths}. Agreement is
found with \citet{sandlin77} and \citet{harper99}.

\subsection{Silicon isoelectronic sequence}

There are four  key forbidden transitions for this sequence which are (in decreasing
wavelength order) $^3P_{2,1}$--$^1D_2$, $^1D_2$--$^1S_0$ and
$^3P_1$--$^1S_0$.

\subsubsection{Cl IV}

The three decays from the $^1D_2$ level to $^3P_J$ lie outside the
wavelength range of UVES, between 7260 and 8050~\AA. The
$^3P_2$--$^1S_0$ transition is blended with the $n=5$ member of the
\ion{He}{ii} Fowler series at 3204~\AA\ and can not be resolved. The
remaining transitions, $^3P_1$--$^1S_0$  and $^1D_2$--$^1S_0$, are
observed in the UVES spectrum and the wavelengths are given in
Table~\ref{tbl.wavelengths}, 
where good agreement is found with the values of \citet{bowen60}.

\subsubsection{Ar V}

The $^3P_2$--$^1D_2$ and $^1D_2$--$^1S_0$ transitions lie outside the
UVES wavelength ranges, at 7005.7 and 4625.5~\AA, respectively. The
weak  $^3P_0$--$^1D_2$ transition (6133.1~\AA) can not be found in the
spectrum, but $^3P_1$--$^1D_2$  is clearly seen and the wavelength is
given in Table~\ref{tbl.wavelengths} where good agreement is found
with the \cite{bowen60} measurement. The $^3P_1$--$^1S_0$ transition
is found in the STIS spectrum at 2691.85~\AA\ but the wavelength is
0.11~\AA\ longer than the IUE measurement of \citet{penston83}, which
is outside the uncertainties of the two measurements. However, the STIS wavelength is
consistent with the \citet{bowen60} wavelength which was derived from
longer wavelength lines. The $^1S_0$ level also decays to $^3P_1$ with
an expected wavelength of 2786.8~\AA, but it is predicted to be about
a factor 100 weaker than \lam2691.85 and it can not be found in the
STIS spectrum.

\subsubsection{K VI}

The $^3P_{1,2}$--$^1D_2$ transitions are both found in the UVES
spectra at 5603.8 and 6230.1~\AA, respectively, and their wavelengths
are consistent with the values of \citet{bowen60} although the new
wavelengths have a significantly improved accuracy. The
$^1D_2$--$^1S_0$ line with expected 
wavelength 4101.6~\AA\ lies between the two wavelength ranges of UVES
and is not observed. 

The UV transition $^3P_1$--$^1S_0$, has not previously been identified
and we believe it is the line at 2368.26~\AA\ in the STIS
spectrum. The \ion{K}{vi} energy levels determined by \citet{smitt76}
from allowed transitions in the extreme ultraviolet yield a predicted
wavelength of 2368.24~\AA\ with an accuracy of around 0.11~\AA, in
excellent agreement with the STIS measurement. Note, however, that the
STIS line is very weak and partly blended with an \ion{Fe}{ii} line at
2367.41~\AA\ thus there is some uncertainty over this identification.

\subsubsection{Ca\,VII}

The $^3P_{1,2}$--$^1D_2$ transitions are well-observed in the UVES
spectrum at 4940.6 and 5620.1~\AA, respectively, and the wavelengths
are in agreement with the values reported for RR Tel by
\citet{thackeray74}. 

The $^3P_1$--$^1S_0$ transition is found at 2111.49~\AA\ and is a
strong line in the STIS spectrum. It was not measured by
\citet{penston83} probably due to the low instrument sensitivity at
this wavelength. \citet{smitt76} derived energy levels for
\ion{Ca}{vii} using allowed transitions at EUV wavelengths, and the two
forbidden line measurements of \citet{thackeray74} combined with these energies
yield a predicted wavelength of 2111.64~\AA, close to the measured
STIS wavelength. \lam2111.5 has previously been measured from a STIS
spectrum of another symbiotic star, AG Draconis, by
\citet{young06}. We note that if the new reference wavelength from RR Tel is
used then the velocity of the AG Dra line becomes $-144$~\kms\ in very
good agreement with other emission lines from that system.

The $^1D_2$--$^1S_0$ transition is blended with a
much stronger transition due to
the $n=19$ member of the \ion{H}{i} Balmer
series (\lam3687.91) and an accurate measurement of the \ion{Ca}{vii} line can not
be made. However combining the measured wavelengths of the
$^3P_1$--$^1S_0$ and $^3P_{1}$--$^1D_2$ transitions yields a predicted
wavelength of 3687.37~\AA\ (vacuum), placing it
on the short wavelength side of the blending \ion{H}{i} line (3687.91~\AA).

\subsection{Phosphorus isoelectronic sequence}

The phosphorus sequence ions have the same ground configuration terms
as nitrogen-like ions, and so yield eight forbidden transitions seven
of which are potentially observable.

\subsubsection{Ar IV}

Only two of the \ion{Ar}{iv} forbidden transitions are found in the
RR Tel spectra: the $^4S_{3/2}$--$^2P_{1/2,3/2}$ transitions at 2869.1
and 2854.6~\AA, respectively. Both lines were reported by
\citet{penston83} from IUE spectra and the STIS measurements agree
with these within the uncertainties. The \citet{bowen60} wavelengths
reported in Table~\ref{tbl.wavelengths} were derived indirectly from
transitions measured at visible wavelengths.

\subsubsection{K V}

The NIST energies for \ion{K}{v} are from \citet{smitt76} who used the
forbidden wavelengths of \citet{bowen60} together with EUV
measurements of allowed transitions to determine the ground
configuration energies. \citet{bowen60} only measured the
$^4S_{3/2}$--$^2D_{3/2,5/2}$ transitions which, at 4163.30 and
4122.63~\AA\ (air), are not available in the UVES spectra. \citet{thackeray77}
measured both of these transitions in optical spectra of RR Tel and
found wavelengths of 4163.55 and 4122.75~\AA\ (air). 

The energies provided by \citet{smitt76} put the
$^4S_{3/2}$--$^2P_{1/2,3/2}$ transitions at 2515.21 and 2495.00~\AA,
respectively. The latter is an excellent wavelength match for a line
in the STIS spectrum at 2494.97~\AA, and the width of the line is also
consistent with lines of similar ionization potential (e.g.,
\ion{O}{iv}). The \lam2515.2 line is expected to be 2--4 times weaker
based on the CHIANTI atomic model, and a good match is the observed
line at 2515.35~\AA, which also has a width consistent with
\lam2494.97. The line is partly blended with a weak \ion{Fe}{ii}
transition and the components have been resolved by fitting two Gaussians.

The $^2D_{3/2,5/2}$--$^2P_{1/2,3/2}$ transitions occur between 6223
and 6351~\AA, with the strongest expected to be 3/2--3/2 at
6223.7~\AA. A line is found at this wavelength in the UVES spectrum
although it is blended with another line in the short wavelength
wing. A gaussian was fit to the \ion{K}{v} line by ignoring the wing
in the fitting process.
The wavelength is given in Table~\ref{tbl.wavelengths} and is in
excellent agreement with that predicted from the 
\citet{smitt76} energy levels.

\subsubsection{Ca VI}

The four $^2D$--$^2P$ transitions are found between 5462 and 5767~\AA, and two of the lines have previously been measured in RR Tel by \citet{thackeray74} \citep[see also][]{thackeray77}: the 3/2--3/2 and 5/2--3/2 transitions at 5460.7 and 5586.3~\AA\ (air wavelengths). Both lines are also found in the UVES spectrum and the wavelengths are in good agreement. Note that the wavelengths of \citet{bowen60} quoted in Table~\ref{tbl.wavelengths} were derived through calculation or interpolation along the isoelectronic sequence are so are of limited accuracy. The 5/2--1/2 transition (5767.0~\AA) is expected to be very weak and is not observed, while the 3/2--1/2 transition (5633.3~\AA) is blended with \ion{Fe}{vi} \lam5632.6. 

The UVES spectra contain the $^4S_{3/2}$--$^2D_{3/2,5/2}$ transitions
at 3670.2 and 3726.5~\AA. The latter is the stronger line and has a weaker
\ion{O}{ii} line in the long wavelength wing. The \ion{Ca}{vi}
centroid was estimated by performing a single Gaussian fit, but
ignoring the line wing. We note that \citet{thackeray74} suggested there was a strong \ion{O}{iv} contribution at this wavelength, but this is not the case in the UVES spectrum. \lam3670.2 is blended with the $n=25$ member 
of the \ion{H}{i} Balmer series which provides the dominant
contribution. An accurate measurement of the line's centroid could not
be made and so the transition is not listed in
Table~\ref{tbl.wavelengths}.

The $^4S_{3/2}$--$^2P_{3/2,1/2}$ transitions are found at 2215.2 and
2242.8~\AA, respectively, in the STIS spectra. Both are strong, well-observed lines,
however \lam2215.2 is blended with the $n=11$ member of the
\ion{He}{ii} Fowler series. Comparisons with other members of the
Fowler series suggest that \ion{He}{ii} contributes around one third
to the blend, while the wavelength of the blended feature is shorter
than expected for the \ion{He}{ii} line, suggesting \ion{Ca}{vi} is on
the short wavelength side of the profile and \ion{He}{ii} on the long
wavelength side. Despite this a two Gaussian fit to the feature fails
to yield a \ion{He}{ii} fit consistent with the expected flux and wavelength. To
estimate the \ion{Ca}{vi} wavelength we fit a single Gaussian to the
blended feature and apply a $-5$~\kms\ shift to the resulting centroid
to indicate  that \ion{Ca}{vi} is on the short wavelength side of the
fitted line. An  uncertainty of 5~\kms\ has been added to the other
sources of error for this centroid to yield the wavelength given in
Table~\ref{tbl.wavelengths}. \citet{penston83} measured both \ion{Ca}{vi} lines in IUE spectra of RR Tel, but did not note the blend of \lam2215.2 with \ion{He}{ii}. This perhaps explains the 4~\kms\ discrepancy for this line; the wavelengths for \lam2242.8 are in good agreement.

A consistency check on the \ion{Ca}{vi} levels can be performed using
the \lam\lam2215.2, 3726.5 and 5462.2 lines. We find  a predicted
wavelength for the $^4S_{3/2}$--$^2P_{1/2}$ transition of $2215.158\pm
0.032$~\AA\ which is in good agreement with the measured
wavelength. We can also use the measured wavelengths for the \lam\lam2215.2 and 5587.8 lines to predict a wavelength for the $^4S_{3/2}$--$^2D_{3/2}$ transition which we find to be $3670.15\pm 0.16$~\AA, placing it on the short wavelength side of the stronger \ion{H}{i} line (3670.51~\AA).

\subsection{Sulphur isoelectronic sequence}

There are six forbidden transitions for sulphur-like ions, but two ($^3P_0$--$^1D_2$ and $^3P_2$--$^1S_0$) are generally  too weak to observe.

\subsubsection{Ar\,III}

The energies for the \ion{Ar}{iii} ground configuration levels in the
NIST database are given to three decimal places, suggesting a very
high accuracy. The levels were derived from unpublished work
so the measurement sources are unknown, although the
wavelengths derived from the level energies suggest they are  based on
the wavelengths of \citet{bowen60} who gave quite
precise wavelengths for the four strong forbidden
transitions. The $^3P_1$--$^1S_0$ and $^1D_2$--$^1S_0$
transitions 
have been reported by \citet{thackeray77} from optical spectra of RR
Tel, and $^3P_1$--$^1S_0$ has been reported by \citet{penston83} from
ultraviolet spectra. The $^3P_{1,2}$--$^1D_2$ transitions at 7751.1
and 7135.8~\AA\ lie outside of the UVES wavelength range and so can
not be measured here.

The $^3P_1$--$^1S_0$ transition at 3110.1~\AA\ is the longest
wavelength line seen in the STIS UV spectrum, and is also one of only
two lines that are also observed by UVES (the other is \ion{Ni}{vii}
\lam3106.2). The STIS line profile is symmetric, while the UVES
profile has an enhanced long wavelength wing. Fitting the UVES profile
with a single Gaussian yields a wavelength within 0.005~\AA\ of the
STIS line. The wavelength given in Table~\ref{tbl.wavelengths} is from
the STIS spectrum, and agrees well with the \citet{bowen60} and NIST
values but 
there is a significant discrepancy with the \citet{thackeray77} and
\citet{penston83} wavelengths.

The $^1D_2$--$^1S_0$  transition at 5193.1~\AA\ has a very extended
long wavelength wing that may partly include another emission
line. This wing has been masked out when performing the fit so that
only the central portion of the profile and the short wavelength wing
are included. The derived wavelength agrees within the uncertainties
with the measurements of \citet{bowen60} and \citet{thackeray77}.

\subsubsection{K\,IV}

The $^1D_2$--$^1S_0$ transition (4512.2~\AA) lies outside of the UVES wavelength ranges and so can not be measured. The strongest transition, $^3P_2$--$^1D_2$, however is clearly identified at 6103.5~\AA\ and the wavelength is in agreement with the measurements of \citet{bowen60} and \citet{thackeray77}. The latter author also identified the $^3P_1$--$^1D_2$ transition at 6792.5~\AA\ (air) in the RR Tel optical spectrum and the same line can be seen in the present UVES spectra. However, \citet{feucht97} provided an accurate measurement of the $^3P_2$--$^3P_1$ splitting which, using the measured \lam6103.5 wavelength, implies a wavelength for $^3P_1$--$^1D_2$ of 6797.04~\AA. A group of four weak lines can be found near this wavelength, one of which has a wavelength of 6797.05~\AA. Therefore we tentatively make this identification.

The $^3P_1$--$^1S_0$ transition is expected in the ultraviolet at 2711.9~\AA, but can not be found in the present STIS spectrum.

\subsubsection{Ca\,V}

Of the four potentially observable \ion{Ca}{v} lines, one line,
\lam3999.0, is outside of the UVES wavelength ranges, while another, \lam6088.1, is blended with a \ion{Fe}{vii} line. \citet{young05a} estimated that the \ion{Ca}{v} line contributes $<$~2~\%\ to this blend and so it can not be used for determining the rest wavelength of the line.

The $^3P_2$--$^1D_2$ transition at 5310.6~\AA\ was previously measured by \citet{bowen60} and \citet{thackeray77} and the present measurement is in good agreement with the latter, but marginally discrepant with the former. In the STIS spectrum the $^3P_1$--$^1S_0$ transition is found at 2413.6~\AA\ and the measured wavelength is in good agreement with the previous value of \citet{penston83}.

\section{Energy levels}\label{sect.energy}

The wavelengths provided in Table~\ref{tbl.wavelengths} can be used to
determine new level energies for the different ions, and these are
given in Tables~\ref{tbl.e-b}--\ref{tbl.e-s} for the different
isoelectronic sequences. In some cases an energy is not determined
directly but requires an additional energy estimate, such as the
$^3P_1$--$^1D_2$ transition in the carbon and oxygen sequences, where
the $^3P_0$--$^3P_1$ energy is required. For most such instances, the
splitting has been obtained to high accuracy at infrared wavelengths,
and the values are included in the tables, but enclosed with parentheses.

The sections below give specific details about how the energies were
derived for each ion.

\subsection{Beryllium isoelectronic sequence}

All five of the intercombination transitions were measured for each of
\ion{C}{ii}, \ion{N}{iii} and \ion{O}{iv}. The energies for the
$^4P_{1/2}$ and $^4P_{3/2}$ levels are obtained directly from the
decays to the ground $^2P_{1/2}$ level, while for $^4P_{5/2}$ it was
necessary to use the energy for the ground $^2P_{3/2}$ level. For
both \ion{C}{ii} and \ion{O}{iv} this energy has been measured
directly at infrared wavelengths to high accuracy, but the authors are
not aware of a similar measurement for \ion{N}{iii} so the
$^2P_{1/2,3/2}$--$^4P_{1/2,3/2}$ transitions from the present work
were used to yield average values of the $^2P$ splitting. Since the
separations of the intercombination lines do not depend on the
absolute wavelength calibration of the spectrum, then only the error
components from the Gaussian fitting and the echelle order comparison
(the \ion{N}{iii} lines are spread across two echelle orders)
were used to derive the error on the energy.

\subsection{Carbon isoelectronic sequence}

For \ion{N}{ii} the $^1S_0$ and $^5S_2$ level energies can be determined from
the measured lines by making use of previously determined energies for
the $^3P_1$ and $^1D_2$ levels. The $^3P_{1}$ level energy has been very
accurately measured in the laboratory by \citet{brown94}, while the
most accurate
$^1D_2$ level energy is from \citet{spyro95} from planetary nebulae
observations. Of the two RR Tel transitions that can potentially be
used to derive the $^1S_0$ level energy, it is the $^1D_2$--$^1S_0$
transition that yields the smallest error bar and so this is used in
Table~\ref{tbl.e-c}.

Very precise measurements of the $^3P_1$ and $^3P_2$ levels  are
available for \ion{Ne}{v} from \citet{feucht97}, and these have been
used to yield the $^1D_2$, $^1S_0$ and $^5S_2$ levels from the
\lam\lam3426.9, 1574.7 and 1145.6 wavelengths, respectively.

The \ion{Na}{vi} $^1D_2$ level energy is obtained from the measured
\lam2872.7 and \lam2971.8 lines by making use of the $^3P_{1,2}$
energies obtained by \citet{feucht97} from infrared spectra. The
energies derived from each of the RR Tel lines were averaged to yield
the energy in Table~\ref{tbl.e-c}. Both \lam1356.3 and \lam2569.6
decay from the $^1S_0$ level and have been used to derive the level's
energy. The $^1D_2$ energy from Table~\ref{tbl.e-c} was used in the
former case. The energies obtained from the two RR Tel lines were
averaged to yield
the energy in Table~\ref{tbl.e-c}. For both $^1D_2$ and $^1S_0$ the
energies derived here are significantly different to the NIST values.

\subsection{Nitrogen isoelectronic sequence}

Since the \ion{Ne}{iv} $^4S$--$^2D$ transitions may arise from the low density
component of the nebula (Sect.~\ref{sect.ne4} and Appendix~\ref{app.model}) we
do not use the measured lines to derive the $^2D$ energies. Instead we
make use of the $^2D$--$^2P$ wavelengths of \citet{bowen60} to derive
the $^2D$ energies from the $^2P$ energies that are determined from
our wavelengths of the \lam1601.5, 1601.7 lines. Note that the $^2D$
and $^2P$ splittings of the measured RR Tel lines are consistent with
those of \citet{bowen60}, but the \citet{bowen60} splittings
themselves are inconsistent with each other within the error bars. A
new measurement of the $^2D$--$^2P$ wavelengths would be valuable.

The \ion{Na}{v} \lam1365.4 and \lam1366.1 lines directly yield energies for the
$^2P_{1/2,3/2}$ levels, but Sect.~\ref{sect.na5} suggested that the
lines from the $^2D$ levels may arise from the low density, redshifted
nebula component. The \citet{bowen60} $^2D$--$^2P$ level splittings
are not very accurate and so can not be used to determine the $^2D$
energies as done for \ion{Ne}{iv}. In Table~\ref{tbl.e-n} we therefore
list the $^2D_{5/2}$ energy derived from the STIS \lam2069.9 line, with a
note that it may be uncertain. Note that all three energies derived in
the present work are significantly different from the NIST energies.

A discrepancy amongst the \ion{Mg}{vi} wavelengths led to the
introduction of an additional error factor in the present wavelength
study and thus some caution needs to be applied in considering the
energies from this ion. Our preferred choice for Table~\ref{tbl.e-n}
is to use the \lam1190.0 and \lam1191.6 lines to yield the $^2P$
energies, and then to derive the $^2D$ energies from these using the
visible $^2D$--$^2P$ transitions. This is principally because, if
there are multiple plasma components with different velocities that
are causing the \ion{Mg}{vi} discrepancy, then they will affect the
transitions from the $^2D$ term differently to those from $^2P$
term. The $^2D$ energies resulting from this method are significantly
different to the NIST energies.

\subsection{Oxygen isoelectronic sequence}

The \ion{Ne}{iii} $^1D_2$ and $^1S_0$ level energies can be obtained from the RR Tel
spectra and are shown in Table~\ref{tbl.e-o}. The $^1D_2$ energy is
obtained directly from \lam3869.8, while the \lam1814.6 wavelength is
combined with the $^3P_1$ energy from \citet{feucht97} to yield the
$^1S_0$ energy. The latter shows a small, but significant discrepancy
with the value in the NIST databae.

\ion{Na}{iv} \lam3242.7 is used to determine the $^1D_2$ level energy,
while the separation of \lam3242.7 and \lam3363.3 is used to yield the
$^3P_1$ energy. Since the line separation does not depend on the
absolute calibration, then the wavelength uncertainties include only
the fitting error and the echelle order comparison error. The
$^3P_2$--$^3P_1$ transition has been measured at infrared wavelengths
by \citet{kelly95} who were able to determine the hyperfine
splitting. The two strongest components are at 1105.88 and
1106.12~cm$^{-1}$ in good agreement with the $^3P_2$--$^3P_1$ energy
found here.

The separation of the two strong \ion{Mg}{v} $^3P_{2,1}$--$^1D_2$ transitions in
the RR Tel spectrum yields the separation of the $^3P_{2,1}$ levels. As for
\ion{Na}{iv}, the only error components to consider in this case are
the fitting errors and echelle order comparison error. The resulting
energy (Table~\ref{tbl.e-o}) is in good agreement with the value of
$1782.58\pm 0.20$
measured directly from infrared spectra by \citet{feucht97}.  The
$^1D_2$ energy is determined directly from the $^3P_{2}$--$^1D_2$
transition, while the $^1S_0$ level is derived from the \lam1324.4
wavelength and the $^3P_1$ energy from Table~\ref{tbl.e-o}. Both
$^1D_2$ and $^1S_0$ show significant discrepancies with the NIST
energies. 

The separation of  \ion{Al}{vi} \lam2430.2 and \lam2603.1 yields the
$^3P_1$ energy (Table~\ref{tbl.e-o}) which is in good agreement with
the value of $2732.46\pm 0.46$ measured from infrared spectra by
\citet{feucht01}. Again, only the fitting errors and echelle order 
comparison error are used to derive the energy uncertainty. The
$^1D_2$ energy is determined directly from the \lam2430.2 wavelength,
and shows a significant discrepancy with the NIST energy.

\subsection{Magnesium isoelectronic sequence}

The \lam1467.4 and \lam1199.2 lines of \ion{P}{iv} and \ion{S}{iv}
directly yield the $3s3p$ $^3P_1$ energies shown in
Table~\ref{tbl.e-mg}.

\subsection{Aluminium isoelectronic sequence}

Only energies for the \ion{S}{iv} levels can be derived from the RR
Tel data-set. The ground term splitting was measured very accurately
by J.H.~Lacy from planetary nebulae infrared spectra (unpublished
work)  and the value of
$951.43\pm 0.01$~cm$^{-3}$ was reported in
\citet{kaufman93}. By considering the splitting of the \lam1398.0 and
\lam1416.9 lines, the ground term splitting is found to be $951.42\pm
1.0$~cm$^{-3}$, in excellent agreement  with the infrared value.

The \ion{S}{iv} $^4P_{1/2}$ energy is obtained from the measured
\lam1423.9 wavelength by adding the $^2P_{3/2}$ energy value. Although
the $^4P_{3/2}$ energy is obtained directly from the \lam1398.0
wavelength, a more accurate value is obtained from the \lam1416.9
wavelength by adding the $^2P_{3/2}$ energy (the \lam1398.0 line
profile is very noisy). Finally, the $^4P_{5/2}$ energy is obtained
from the \lam1406.0 measured wavelength by also adding the $^2P_{3/2}$ energy.

\subsection{Silicon isoelectronic sequence}

The \ion{Cl}{iv} \lam3119.5 and \lam5324.7 wavelengths allow the
$^1S_0$ and $^1D_2$ level energies to be determined, respectively. The
former line decays
to $^3P_1$ rather than the ground level, but this level energy was
accurately measured from infrared spectra by \citet{feucht01}. With
the $^1S_0$ energy determined, the \lam5324.7 wavelength then yields
the $^1D_2$ energy.

The \ion{Ar}{v} \lam2691.8 and \lam6436.9 lines both decay to $^3P_1$
and yield the $^1S_0$ and $^1D_2$ level energies, respectively, when
the $^3P_1$ energy from the \citet{feucht97} is used.

The same transitions were measured for \ion{K}{vi}, and give the
$^1D_2$ and $^1S_0$ level energies shown in Table~\ref{tbl.e-si} when
combined with the $^3P_1$ level energy measured by \citet{kelly95}
from infrared spectra. The additional $^3P_2$--$^1D_2$ transition
measured at 6230.1~\AA\ in the RR Tel spectra also allows the $^3P_2$
energy to be determined. We use the separation of the \lam5603.8 and
\lam6230.1 lines to give the $^3P_1$--$^3P_2$ separation to which is
then added the \citet{kelly95} $^3P_1$ energy. Note that, since the
\lam5603.8, \lam6230.1 separation does not depend on the absolute
wavelength calibration, then the only error sources considered were
the line fitting errors and the echelle order comparison
uncertainty. The $^3P_2$ energy that results shows a small, but
significant difference with the NIST energy.

The measured \ion{Ca}{vii} \lam2111.5 and \lam4940.6 wavelengths yield the energies
of the $^1S_0$ and $^1D_2$ energies, respectively, when combined with
the energy for the $^3P_1$ level. The latter has not been measured
from infrared spectra and so we use the value of $1624.9\pm
1.1$~cm$^{-1}$ from \citet{smitt76}. The separation of the \lam4940.6
and \lam5620.1 lines yields a value 
of the $^3P_1$--$^3P_2$ energy of $2447.11\pm 0.14$~cm$^{-1}$ that is more accurate than the direct
measurement of the transition
by \citet{feucht01} of $2447.5\pm 0.7$~cm$^{-1}$. Since the absolute
calibration is not important for determining the separation of
emission lines, the only error contributions considered for the RR Tel
lines were the fitting error and echelle order comparison
uncertainty. Combining the
$^3P_1$--$^3P_2$ separation with the $^3P_1$ energy yields the $^3P_2$ energy
shown in Table~\ref{tbl.e-si}.

\subsection{Phosphorus isoelectronic sequence}

The $^2P_{1/2,3/2}$ energies of \ion{Ar}{iv} were directly obtained
from the \lam2869.1 and \lam2854.6 wavelengths respectively.

The \ion{K}{v} $^2P_{1/2,3/2}$ level energies are  obtained
directly from the measured \lam2869.1 and \lam2854.6 lines,
respectively. The $^2P_{3/2}$ energy is then combined with the
measured $^2D_{3/2}$--$^2P_{3/2}$ wavelength to yield the $^2D_{3/2}$
energy shown in Table~\ref{tbl.e-p}.

All of the \ion{Ca}{vi} ground configuration energies can be
determined from the RR Tel spectra, and $^2P_{1/2,3/2}$ and
$^2D_{3/2}$ are each determined directly from the measured wavelengths
of \lam2242.7, \lam2215.2 and \lam3726.4, respectively. The
$^2D_{5/2}$ energy is derived from the wavelength of the
$^2D_{5/2}$--$^2P_{3/2}$ transition, by making use of the $^2P_{3/2}$
energy. Note that the `x' in the NIST column energies denotes that 
there is a potential systematic offset for the ground configuration
energies due to how the energies were derived. The energy values
presented here remove this uncertainty.

\subsection{Sulphur isoelectronic sequence}

The \ion{Ar}{iii} \lam3110.1 and \lam5193.1 measured wavelengths yield
the $^1S_0$ and $^1D_2$ level energies when combined with the $^3P_1$
energy that was directly measured from infrared spectra by
\citet{kelly95}.

The \ion{K}{iv} $^1D_2$ level energy is obtained directly from the
measured \lam6103.5 wavelength.

The \ion{Ca}{v} $^1D_2$ and $^1S_0$ level energies are obtained from
the measured wavelengths of \lam5310.8 and \lam2413.6,
respectively. For the former, the $^3P_1$ energy of \citet{feucht01}
is required to yield the energy shown in Table~\ref{tbl.e-s}. 

\begin{deluxetable}{llccccccccccc}
\tabletypesize{\footnotesize}
\tablecaption{Level energies for boron-like ions.\label{tbl.e-b}}
\tablehead{
  &&  \multicolumn{11}{c}{Energies / cm$^{-1}$} \\
  \cline{3-13}
  &&
  \multicolumn{3}{c}{\ion{C}{ii}} &&
  \multicolumn{3}{c}{\ion{N}{iii}} &&
  \multicolumn{3}{c}{\ion{O}{iv}} \\
  \cline{3-5}\cline{7-9}\cline{11-13}
  \colhead{Configuration} &
  \colhead{Level} &
  \multicolumn{2}{c}{RR Tel} &
  NIST &&
  \multicolumn{2}{c}{RR Tel} &
  NIST &&
  \multicolumn{2}{c}{RR Tel} &
  NIST 

}
\startdata
$2s^22p$ &$^2P_{1/2}$&0.0&&0.0 &&0.0&&0.0&&0.0&&0.0\\
&$^2P_{3/2}$&(63.397)\tablenotemark{a}&\nodata&63.4 &
           &173.96&0.43 &174.4&
           &(386.245)\tablenotemark{b}&\nodata&385.9\\
\noalign{\smallskip}
$2s2p^2$&$^4P_{1/2}$&43003.2&0.9&43003.3 &
                   &57187.4&1.1&57187.1&
                   &71440.5&1.5&71439.8\\
&$^4P_{3/2}$&43024.2&0.9&43025.3 &
           &57247.0&1.2&57246.8&
           &71571.8&1.5&71570.1\\
&$^4P_{5/2}$&43053.4&0.8 &43053.6 &
           &57327.8&1.2&57327.9&
           &71755.8&1.5&71755.5\\
\enddata
\tablenotetext{a}{\citet{cooksy86}}
\tablenotetext{b}{\citet{feucht97}}
\end{deluxetable}

\begin{deluxetable}{llccccccccccccc}
\tabletypesize{\scriptsize}
\tablecaption{Level energies for carbon-like ions.\label{tbl.e-c}}
\tablehead{
  &&  \multicolumn{11}{c}{Energies / cm$^{-1}$} \\
  \cline{3-13}
  &&
  \multicolumn{3}{c}{\ion{N}{ii}} &&
  \multicolumn{3}{c}{\ion{Ne}{v}} &&
  \multicolumn{3}{c}{\ion{Na}{vi}} \\
  \cline{3-5}\cline{7-9}\cline{11-13}
  \colhead{Configuration} &
  \colhead{Level} &
  \multicolumn{2}{c}{RR Tel} &
  NIST &&
  \multicolumn{2}{c}{RR Tel} &
  NIST &&
  \multicolumn{2}{c}{RR Tel} &
  NIST 
}
\startdata
$2s^22p^2$&$^3P_0$&0.0&&0.0 &&0.0&&0.0&&0&&0\\
&$^3P_1$ &(48.738)\tablenotemark{a}&\nodata&48.7 &
         &(411.226)\tablenotemark{c}&\nodata&411.227&
         &(694.62)\tablenotemark{c}&\nodata &694.62\\
&$^3P_2$ &(130.774)\tablenotemark{a}&\nodata&130.8 &
         &(1109.468)\tablenotemark{c}&\nodata&1109.467&
         &(1855.98)\tablenotemark{c}&\nodata&1855.98\\

&$^1D_2$ &(15316.17)\tablenotemark{b}&\nodata&15316.2 &
         &30290.3&0.6&30290.67&
         &35505.7&0.7 &35498\\

&$^1S_0$ &32688.04 &0.35 &32688.8 &
         &63916.6 &1.3 &63915.4&
         &74423.0 &1.9 &74414\\
\noalign{\smallskip}
$2s2p^3$ &$^5S_2$&46784.6&1.0&46784.6 &&88400.7&2.0&88399.5&
        &\nodata&\nodata&103010+x\\
\enddata
\tablenotetext{a}{\citet{brown94}}
\tablenotetext{b}{\citet{spyro95}}
\tablenotetext{c}{\citet{feucht97}}
\end{deluxetable}

\begin{deluxetable}{lccccccccccccc}
\tabletypesize{\footnotesize}
\tablecaption{$2s^22p^3$ level energies for nitrogen-like ions.\label{tbl.e-n}}
\tablehead{
  &  \multicolumn{11}{c}{Energies / cm$^{-1}$} \\
  \cline{2-12}
  &
  \multicolumn{3}{c}{\ion{Ne}{iv}} &&
  \multicolumn{3}{c}{\ion{Na}{v}} &&
  \multicolumn{3}{c}{\ion{Mg}{vi}} \\
  \cline{2-4}\cline{6-8}\cline{10-12}
  \colhead{Level} &
  \multicolumn{2}{c}{RR Tel} &
  NIST &&
  \multicolumn{2}{c}{RR Tel} &
  NIST &&
  \multicolumn{2}{c}{RR Tel} &
  NIST 
}
\startdata
$^4S_{3/2}$&0.0&&0.0&&0.0&&0.0&&0.0&&0.0\\
$^2D_{5/2}$ &41228.4&1.3 &41234.1+e&
            &48311.1\tablenotemark{a}&1.4 &48343.0&
            &55361.9&1.8&55356.0\\
$^2D_{3/2}$ &41273.1&1.3 &41279.5+e&
            &\nodata&\nodata&48379.0&
            &55376.4&1.8&55372.8\\
$^2P_{1/2}$ &62433.7& 1.3& 62434.6+e&
            &73202.1&1.5&73232.7&
            &83921.6&1.7&83920.0\\
$^2P_{3/2}$ &62441.4 &1.3&62441.3+e&&73239.3&1.5&73267.1&
            &84030.8&1.7&84028.4\\
\enddata
\tablenotetext{a}{Derived from a line that may be redshifted relative
  to other lines in the RR Tel nebula.}
\end{deluxetable}

\begin{deluxetable}{lccccccccccccccccc}
\tabletypesize{\scriptsize}
\tablecaption{$2s^22p^4$ level energies for oxygen-like ions.\label{tbl.e-o}}
\tablehead{
  &  \multicolumn{15}{c}{Energies / cm$^{-1}$} \\
  \cline{2-16}
  &
  \multicolumn{3}{c}{\ion{Ne}{iii}} &&
  \multicolumn{3}{c}{\ion{Na}{iv}} &&
  \multicolumn{3}{c}{\ion{Mg}{v}} &&
  \multicolumn{3}{c}{\ion{Al}{vi}} \\
  \cline{2-4}\cline{6-8}\cline{10-12}\cline{14-16}
  \colhead{Level} &
  \multicolumn{2}{c}{RR Tel} &
  NIST &&
  \multicolumn{2}{c}{RR Tel} &
  NIST &&
  \multicolumn{2}{c}{RR Tel} &
  NIST &&
  \multicolumn{2}{c}{RR Tel} &
  NIST 
}
\startdata
$^3P_2$&0.0&&0.0 &&0.0&&0.0&&0.0&&0.0&&0&&0\\
$^3P_1$&(642.878)\tablenotemark{a}&\nodata&642.876 &
        &1105.82&0.26&1106.3&
        &1782.68&0.35 &1783.1&
        &2732.67&0.41&2732\\
$^3P_0$&\nodata&\nodata&920.550 &
        &\nodata&\nodata &1576.0&
        &\nodata&\nodata&2521.8&
        &\nodata&\nodata &3829\\
$^1D_2$&25840.8&0.5&25840.72 &
        &30838.9&0.6&30839.8&
        &35924.1&0.7&35926.0&
        &41148.1&0.8&41167\\
$^1S_0$&55750.1&1.1&55752.7 &
        &\nodata&\nodata &66496.0&
        &77286.6&1.6&77279.0&
        &\nodata&\nodata &88213\\
\enddata
\tablenotetext{a}{\citet{feucht97}}
\end{deluxetable}

\begin{deluxetable}{llccccccc}
\tabletypesize{\footnotesize}
\tablecaption{Level energies for magnesium-like ions.\label{tbl.e-mg}}
\tablehead{
  &&  \multicolumn{7}{c}{Energies / cm$^{-1}$} \\
  \cline{3-9}
  &&
  \multicolumn{3}{c}{\ion{P}{iv}} &&
  \multicolumn{3}{c}{\ion{S}{v}} \\
  \cline{3-5}\cline{7-9}
  \colhead{Configuration} &
  \colhead{Level} &
  \multicolumn{2}{c}{RR Tel} &
  NIST &&
  \multicolumn{2}{c}{RR Tel} &
  NIST 
}
\startdata
$3s^2$ &$^1S_0$&0.0&&0.0 &&0.0&&0.0\\
\noalign{\smallskip}
$3s3p$&$^3P_0$&\nodata&\nodata&67918.03 &
              &\nodata&\nodata&83024.0\\
&$^3P_1$&68146.2&1.4&68146.48 &&83391.6&1.7&83393.5\\
&$^3P_2$&\nodata&\nodata&68615.17 &&\nodata&\nodata&84155.2\\
\enddata
\end{deluxetable}

\begin{deluxetable}{llcccc}
\tabletypesize{\footnotesize}
\tablecaption{Level energies for aluminium-like ions.\label{tbl.e-al}}
\tablehead{
  &&  \multicolumn{3}{c}{Energies / cm$^{-1}$} \\
  \cline{3-6}
  &&
  \multicolumn{3}{c}{\ion{S}{iv}} \\
  \cline{3-5}
  \colhead{Configuration} &
  \colhead{Level} &
  \multicolumn{2}{c}{RR Tel} &
  NIST 
}
\startdata
$3s^23p$ &$^2P_{1/2}$&0.0&&0.0 \\
&$^2P_{3/2}$&(951.43)\tablenotemark{a}&\nodata &951.43 \\
\noalign{\smallskip}
$3s3p^2$&$^4P_{1/2}$&71183.2&1.5&71184.1\\
&$^4P_{3/2}$&71527.4&1.4 &71528.7\\
&$^4P_{5/2}$&72073.0 & 1.5 &72074.4\\
\enddata
\tablenotetext{a}{\citet{kaufman93}.}
\end{deluxetable}

\begin{deluxetable}{lccccccccccccccccc}
\tabletypesize{\scriptsize}
\tablecaption{$3s^23p^2$ level energies for silicon-like ions.\label{tbl.e-si}}
\tablehead{
  &  \multicolumn{15}{c}{Energies / cm$^{-1}$} \\
  \cline{3-16}
  &
  \multicolumn{3}{c}{\ion{Cl}{iv}} &&
  \multicolumn{3}{c}{\ion{Ar}{v}} &&
  \multicolumn{3}{c}{\ion{K}{vi}} &&
  \multicolumn{3}{c}{\ion{Ca}{vii}} \\
  \cline{2-4}\cline{6-8}\cline{10-12}\cline{14-16}
  \colhead{Level} &
  \multicolumn{2}{c}{RR Tel} &
  NIST &&
  \multicolumn{2}{c}{RR Tel} &
  NIST &&
  \multicolumn{2}{c}{RR Tel} &
  NIST &&
  \multicolumn{2}{c}{RR Tel} &
  NIST 
}
\startdata
$^3P_0$&0.0&&0.0 &&0.0&&0.0&&0.0&&0.0&&0.0&&0.0\\
$^3P_1$ &(492.351)\tablenotemark{a} &\nodata&492.0 &
         &(763.231)\tablenotemark{b} &\nodata &765.23&
         &(1132.52)\tablenotemark{c} &\nodata &1132.5&
         &(1624.9)\tablenotemark{d}&\nodata&1624.9\\
$^3P_2$ &\nodata &\nodata&1341.9 &
         &\nodata&\nodata&2828.80&
         &2926.44 &0.15 &2927.2&
         &4072.0&1.1&4071.4\\
$^1D_2$&13767.9&0.8&13767.6 &
        &16298.5&0.3&16298.9&
        &18977.5&0.4&18977.8&
        &21865.2&1.2&21864.0\\
$^1S_0$&32548.3&0.6&32547.8 &
        &37912.4&0.8&37912.0&
        &43357.5&2.3&43358.8&
        &48984.9&1.4&48981.4\\
\enddata
\tablenotetext{a}{\citet{feucht01}.}
\tablenotetext{b}{\citet{feucht97}.}
\tablenotetext{c}{\citet{kelly95}.}
\tablenotetext{d}{\citet{smitt76}.}
\end{deluxetable}

\begin{deluxetable}{lccccccccccc}
\tabletypesize{\footnotesize}
\tablecaption{$3s^23p^3$ level energies for phosphorus-like ions.\label{tbl.e-p}}
\tablehead{
  &  \multicolumn{11}{c}{Energies / cm$^{-1}$} \\
  \cline{2-12}
  &
  \multicolumn{3}{c}{\ion{Ar}{iv}} &&
  \multicolumn{3}{c}{\ion{K}{v}} &&
  \multicolumn{3}{c}{\ion{Ca}{vi}} \\
  \cline{2-4}\cline{6-8}\cline{10-12}
  \colhead{Level} &
  \multicolumn{2}{c}{RR Tel} &
  NIST &&
  \multicolumn{2}{c}{RR Tel} &
  NIST &&
  \multicolumn{2}{c}{RR Tel} &
  NIST 

}
\startdata
$^4S_{3/2}$&0.0&&0.0&
                     &0.0&&0.0&
                     &0.0&&0.0\\
$^2D_{3/2}$  &\nodata&\nodata&21090.4 &
             &24013.0&1.1&24012.5&
             &26835.8&0.5&26835.1+x\\
$^2D_{5/2}$  &\nodata&\nodata&21219.3 &
             &\nodata&\nodata&24249.6&
             &27246.9&1.2&27246.6+x\\
$^2P_{1/2}$&34854.3&0.7&34855.5 &
           &39755.9&2.2&39758.1&
           &44589.0&0.9&44586.7+x\\
$^2P_{3/2}$&35031.4&0.7&35032.6 &
           &40080.7&1.0&40080.2&
           &45143.1&1.2&45142.7+x\\
\enddata
\end{deluxetable}

\begin{deluxetable}{lccccccccccc}
\tabletypesize{\footnotesize}
\tablecaption{$3s^23p^4$ level energies for sulphur-like ions.\label{tbl.e-s}}
\tablehead{
  &  \multicolumn{11}{c}{Energies / cm$^{-1}$} \\
  \cline{2-12}
  &
  \multicolumn{3}{c}{\ion{Ar}{iii}} &&
  \multicolumn{3}{c}{\ion{K}{iv}} &&
  \multicolumn{3}{c}{\ion{Ca}{v}} \\
  \cline{2-4}\cline{6-8}\cline{10-12}
  \colhead{Level} &
  \multicolumn{2}{c}{RR Tel} &
  NIST &&
  \multicolumn{2}{c}{RR Tel} &
  NIST &&
  \multicolumn{2}{c}{RR Tel} &
  NIST 

}
\startdata
$^3P_2$  &0.0&&0.0&
         &0.0&&0.0&
         &0.0&&0.0\\
$^3P_1$  &(1112.176)\tablenotemark{a}&\nodata &1112.175&
         &\nodata&\nodata &1671.7&
         &(2404.21)\tablenotemark{b}&\nodata&2404.7\\
$^3P_0$ &\nodata&\nodata&1570.229&
        &\nodata&\nodata&2321.2&
        &\nodata&\nodata&3275.6\\
$^1D_2$&14009.7&0.8&14010.004&
       &16384.0&0.3&16384.0&
       &18829.6&0.4&18830.3\\
$^1S_0$ &33265.8&0.7&33265.724&
       &\nodata&\nodata&33546.3&
       &43836.1&0.9&43836.5\\
\enddata
\tablenotetext{a}{\citet{kelly95}.}
\tablenotetext{b}{\citet{feucht01}.}
\end{deluxetable}

\section{Conclusions}

Ultraviolet and optical spectra of the symbiotic nova RR Telescopii
have been used to derive new and/or updated rest wavelengths for many
forbidden and intercombination transitions of one to six-times ionized
species of ions from C, N, O, F, Ne, Na, Mg, Al, Si, P, S, Cl, Ar and
Ca. The wavelengths have then been used to derive new sets of energy
levels for these ions.

RR Tel is perhaps the best astronomical object for studying forbidden
lines due to the extraordinary brightness of the emission lines, the
ionization structure and density of the nebula,
and the low interstellar extinction along the line-of-sight. Two
complications to the analysis were presented here: the presence of a
low density, redshifted plasma component that contributes and even
dominates the emission of certain lines; and the problem of
determining a rest wavelength scale. For the latter, consistency
checks performed by considering multiple decay paths within certain ions led
to the introduction of an error term that was the largest of those
considered. The most likely source of this error is the presence of
multiple plasma components with different densities and different
velocities, that could lead to the line profiles from transitions with
different excitation potentials to have different
centroids. Future observations with higher spectral resolution and
signal-to-noise in the
ultraviolet would be valuable for studying this further, while
simultaneous ultraviolet and visible observations would be important
for ruling out temporal changes in the line profile shapes (the STIS
and UVES spectra were taken a year apart).

Although the present spectra contain many of the RR Tel forbidden
lines, there crucially exists a gap in the UVES wavelength coverage
from 3914 to 4730~\AA\ that prevents the full complement of forbidden
lines for several ions from being observed. In this regard the present
work is incomplete but with the recent repair of the HST/STIS
instrument there are hopes  that complete, high resolution ultraviolet
and visible
spectra of RR Tel can be obtained in the future.

\acknowledgements
The work of PRY and UF was performed under contract with the Naval Research
Laboratory and was funded by NASA.

{\it Facilities:} \facility{HST(STIS)}, \facility{ESO(UVES)}

\appendix

\section{Modeling contributions to emission lines}\label{app.model}

The standard picture of symbiotic star nebulae is of a large volume of
plasma around the ionization source, with high densities and high
ionization stages close to the source and low densities and low
ionization species far from the source. The \ion{O}{iii} \lam4363 and
\lam5007 line observations presented by \citet{schild97} were
interpreted by the authors as coming from a high density
($\log\,N_{\rm e}> 8$) component at the rest velocity of the system
and a low density ($\log\,N_{\rm e}< 5.5$) component at
$-20$~\kms. The volume of the low density component is $~1000$ times
larger than the high density component. This suggests a geometry of
the nebula with a small, dense component close to the ionization
source and an extended, low density component further from the source.

The velocity of the \ion{O}{ii} \lam3729.9 line presented in this work
suggests that it comes almost entirely from the low density component
on account of the line's velocity shift. In addition the
\lam3727.1/\lam3729.9 density sensitive ratio suggests a density of
$<10^3$~cm$^{-3}$.

The ionization potentials of O$^+$ and O$^{+2}$ are 13.6 and 35.1~eV,
respectively, and one puzzle is why the \ion{Fe}{ii} forbidden lines
have a velocity consistent with formation in the hot, dense component
of the nebula when the ionization potential of Fe$^{+}$ is only 7.9~eV
suggesting the lines should be formed in the extended low density
component. The solution lies in the different sensitivities to density
of the oxygen and iron lines.

We consider a model of the nebula with two densities of $10^4$ and
$10^8$~cm$^{-3}$, and volumes of $V_1$ and $V_2$. Using the CHIANTI
database, emission line fluxes of the two components can be
computed. We assume the same temperature applies to the two plasma
components. The volumes are adjusted to put the \ion{Fe}{ii}
\lam5263.1 components into the ratio of 9:1, which is the approximate
ratio of the two observed components to this line
(Fig.~\ref{fig.uves.cpts}). The low density volume is found to be a factor
$1.5\times 10^6$ larger than the high density volume in this
case. With this volume ratio, the intensities of other emission lines
can be calculated and are shown in Table~\ref{tbl.model}. It is found
that \ion{O}{ii} \lam3729.1 arises almost entirely from the low
density plasma, while \ion{O}{iii} \lam5008.2 has significant
contributions from both plasmas. This demonstrates that, even though
\ion{Fe}{ii} is a low ionization species, it can be principally formed
in the high density plasma. That this is actually happening in RR Tel
is demonstrated by the agreement between the velocities of the
\ion{Fe}{ii} forbidden lines and the high ionization \ion{O}{iv} and
\ion{O}{v} recombination lines (Fig.~\ref{fig.oxy.rec}).

\begin{figure}[h]
\epsscale{0.7}
\plotone{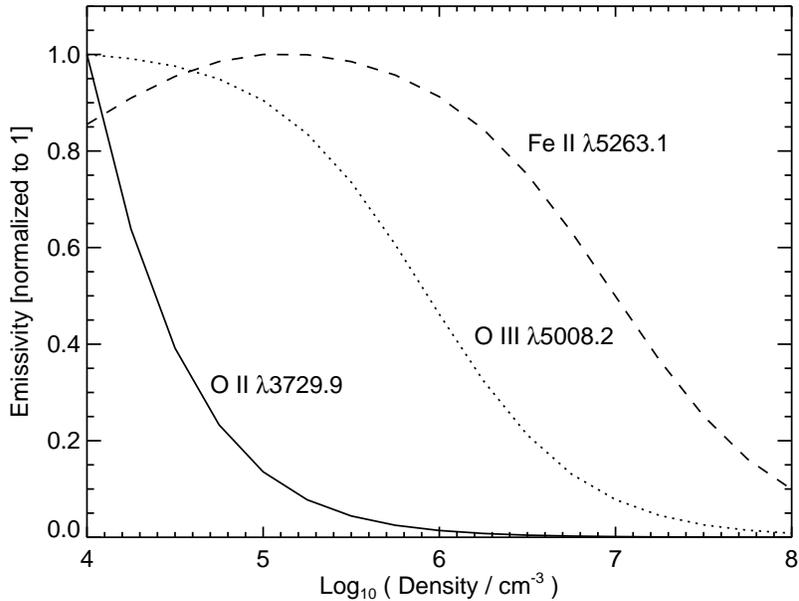}
\caption{Emissivity variation with density for \ion{Fe}{ii}
  \lam5263.1, \ion{O}{ii} \lam3729.9 and \ion{O}{iii} \lam5008.2. Data
calculated from the CHIANTI database.}
\label{fig.emiss}
\end{figure}

The CHIANTI atomic models used in this calculation assume excitation
by electron collisions followed by radiative decay. Excitation through
recombination is not included but should be small for the forbidden
lines considered. The different behaviors of the oxygen and iron lines
are due to how the lines' emissivities change with density, as shown
in Fig.~\ref{fig.emiss}. The quantity plotted is $n_jA_{ji}/N_{\rm e}$
where $n_j$ is the population of the upper level of the atomic
transition, and $A_{ji}$ is the radiative decay rate. This is
proportional to the emissivity of the line, and the results
demonstrate that the \ion{Fe}{ii} line is sensitive to higher
densities than either the \ion{O}{ii} or \ion{O}{iii} lines. The
\ion{O}{ii} line shows very little sensitivity to densities above
$10^6$~cm$^{-3}$, hence it is formed almost entirely in the low
density plasma component.

In general, it is only lines with low excitation potentials that show
significant contributions from the low density component. This is
illustrated by the \lam1601.4 and \lam2425.0 lines of
\ion{Ne}{iv}. The former is a $^4S_{3/2}$--$^2P_{3/2}$ transition with
excitation potential 7.7~eV, while the latter is a
$^4S_{3/2}$--$^2D_{5/2}$ transition with excitation potential 5.1~eV.

Considering other lines listed in Table~\ref{tbl.model}, it can be
seen that the forbidden lines of the beryllium-like ions \ion{C}{iii}
and \ion{N}{iv} are predicted to be mostly formed in the low density
plasma component. This is consistent with the measured wavelengths of
the two lines (Sects.~\ref{sect.c3} and \ref{sect.n4}). The \lam3729.9
line of \ion{O}{ii} corresponds to the atomic transtion
$^4S_{3/2}$--$^2D_{5/2}$ and this same transition for \ion{Ne}{iv} and
\ion{Na}{v} is also predicted to come mostly from the low density
plasma component. There is some evidence from the present spectrum
that the \ion{Ne}{iv} line is redshifted (Sect.~\ref{sect.ne4}) but
this is not clear. Note the \ion{Mg}{vi} $^4S_{3/2}$--$^2D_{5/2}$
transition  is
predicted to come mainly from the high density component in contrast
to the lower ionization stages.

Some caution in interpreting the results shown in
Table~\ref{tbl.model} should be applied as no consideration of the
ionization structure of the nebula is made. The low density model is
likely to be far from the ionization source in reality, and so a high
ionization species such as \ion{Na}{v} may not actually be present in
the low density plasma.

\begin{deluxetable}{lllll}
\tablecaption{\label{tbl.model}}
\tablehead{
  \colhead{Ion} &
  \colhead{Wavelength} &
  \colhead{Transition} &
  \multicolumn{2}{c}{\%\ contribution to line flux} \\
 \cline{4-5}
  &(\AA) &type & Low density & High density 
}
\startdata
\ion{C}{ii}  & 2325.4 & I & 1.2 & 98.8 \\
\ion{C}{iii} & 1906.7 & F & 92.8 & 7.2 \\
\ion{C}{iii} & 1908.7 & I & 0.8 & 99.2 \\
\ion{O}{ii}  & 3729.1 & F & 99.1 & 0.9 \\
\ion{O}{ii}  & 2471.1 & F & 13.3 & 86.7 \\
\ion{O}{iii} & 5008.2 & F & 65.4 & 34.7 \\
\ion{Fe}{ii} & 5263.1 & F & 10.9 & 89.1 \\
\ion{Fe}{ii} & 2365.6 & A & 3.8 & 96.2 \\
\ion{N}{ii}  & 5756.2 & F & 4.8 & 95.2 \\
\ion{N}{ii}  & 6585.3 & F & 93.5 & 6.5 \\
\ion{N}{iv}  & 1483.3 & F & 84.1 & 15.9 \\
\ion{Ne}{iii} & 3869.8 & F & 11.0 & 89.1 \\
\ion{Ne}{iv} & 1601.4 & F & 0.8 & 99.2 \\
\ion{Ne}{iv} & 2425.0 & F & 94.7 & 5.3 \\
\ion{Ne}{v} & 3347.0 & F & 7.9 & 92.1 \\
\ion{Na}{v} & 2067.9 & F & 77.4 & 22.6 \\
\ion{Na}{v} & 1365.4 & F & 0.7 & 99.3 \\
\ion{Mg}{vi} & 1806.0 & F & 9.1 & 90.9 \\
\enddata
\end{deluxetable}

\begin{thebibliography}{}

\bibitem[Bowen(1960)]{bowen60}
  Bowen, I. S. 1960,
  ApJ, 132, 1



\bibitem[Bockasten \& Johansson(1968)]{bockasten68}
  Bockasten, K., \& Johansson, K. B. 1968,
  Ark. Fys., 38, 563

\bibitem[Bromander(1969)]{bromander69}
  Bromander, J. 1969,
  Ark. Fys., 40, 257

\bibitem[Brown et al.(1994)]{brown94}
  Brown, J. M., Varberg, T. D., Evenson, K. M., \& Cooksy, A. L. 1994,
  ApJ, 428, L37

\bibitem[Cooksy et al.(1986)]{cooksy86}
  Cooksy, A. L., Blake, G. A., \& Saykally, R. J. 1986,
  ApJ, 305, L89

\bibitem[Crawford et al.(1999)]{crawford99}
  Crawford, F. L., McKenna, F. C., Keenan, F. P., et al. 1999,
  A\&AS, 139, 135

\bibitem[Curdt et al.(2001)]{curdt01}
  Curdt, W., Brekke, P., Feldman, U., et al. 2001,
  A\&A, 375, 591

\bibitem[Curdt et al.(2004)]{curdt04}
  Curdt, W., Landi, E., \& Feldman, U. 2004,
  A\&A, 427, 1045

\bibitem[Dekker et al.(2000)]{dekker00}
  Dekker, H., D'Odorico, S., Kaufer, A., Delabre, B. \& Kotzlowski,
  H. 2000,
  SPIE, 4008, 534

\bibitem[Dere et al.(1997)]{dere97}
   Dere, K. P., Landi, E., Mason, H. E., Monsignori-Fossi, B. C., \& 
   Young, P. R. 1997,
   A\&AS, 125, 149

\bibitem[Dere et al.(2009)]{dere09}
  Dere K. P., Landi E., Young P. R., et al. 2009,
  \aap, 498, 915

\bibitem[Doschek et al.(1976a)]{doschek76a}
  Doschek, G. A., Bohlin, J. D., \& Feldman, U. 1976a,
  ApJ, 205, L177

\bibitem[Doschek et al.(1976b)]{doschek76}
  Doschek, G. A., Feldman, U., VanHoosier, M. E. \& Bartoe, J.-D. F. 1976b,
  ApJS, 31, 417

\bibitem[Doschek et al.(1977)]{doschek77}
  Doschek, G. A., Feldman, U. \& Cohen, L. 1977,
  ApJS, 33, 101


\bibitem[Edl\'en(1972)]{edlen72}
  Edl\'en, B. 1972,
  Sol. Phys., 24, 356



\bibitem[Feldman \& Doschek(2007)]{feldman07}
  Feldman, U., \& Doschek, G. A. 2007,
  ADNDT, 93, 779

\bibitem[Feuchtgruber et al.(1997)]{feucht97}
  Feuchtgruber, H., Lutz, D., Beintema, D. A., et al. 1997,
  \apj, 487, 962

\bibitem[Feuchtgruber et al.(2001)]{feucht01}
  Feuchtgruber, H., Lutz, D., \& Beintema, D. A. 2001,
  \apjs, 136, 221

\bibitem[Fuhr \& Wiese(2006)]{fuhr06}
  Fuhr, J. R., \& Wiese, W. L. 2006,
  J. Phys. Chem. Ref. Data, 35, 1669


\bibitem[Haisch et al.(1977)]{haisch77}
  Haisch, B. M., Linsky, J. L., Weinstein, A., \& Shine, R. A. 1977,
  ApJ, 214, 785

\bibitem[Harper et al.(1999)]{harper99}
  Harper, G. M., Jordan, C., Judge, P. G., et al. 1999,
  MNRAS, 303, L41

\bibitem[Hartman \& Johansson(2000)]{hartman00}
  Hartman, H. \& Johansson, S. 2000,
  A\&A, 359, 627



\bibitem[Johansson(1983)]{johansson83}
  Johansson, S. 1983,
  MNRAS, 205, 71P



\bibitem[Johansson \& Carpenter(1988)]{johansson88b}
  Johansson, S., Carpenter, K. G. 1988,
  ESA SP-281, Vol. 1, 361


\bibitem[Jordan \& Harper(1998)]{jordan98}
  Jordan, C., \& Harper, G. M. 1998,
  Cool Stars, Stellar Systems and Sun, ASP Conf. Ser. 154
  (eds. R.A. Donahue \& J.A. Bookbinder), CD-1277

\bibitem[Jordan et al.(1994)]{jordan94}
  Jordan, S., M\"urset, U., \& Werner, K. 1994,
  A\&A, 283, 475

\bibitem[Kaufman \& Martin(1993)]{kaufman93}
  Kaufman, V., \& Martin, W. C. 1993,
  J. Phys. Chem. Ref. Data, 22, 279

\bibitem[Keenan et al.(2002)]{keenan02}
  Keenan, F. P., Ahmed, S., Brage, T., et al. 2002,
  MNRAS, 337, 901

\bibitem[Kelly \& Lacy(1995)]{kelly95}
  Kelly, D. M., \& Lacy, J. H. 1995,
  \apj, 454, L161

\bibitem[Kotnik-Karuza et al.(2006)]{kotnik06}
  Kotnik-Karuza, D., Friedjung, M., Whitelock, P. A., et al. 2006,
  A\&A, 452, 503

\bibitem[Kotnik-Karuza et al.(2009)]{kotnik09}
  Kotnik-Karuza, D., Friedjung, M. \& Exter, K. 2009,
  PASJ, 61, 147

\bibitem[Kramida et al.(1999)]{kramida99}
  Kramida, A. E., Bastin, T., Bi\'emont, E., Dumont, P.-D., \& Garnir,
  H.-P. 1999,
  Eur. Phys. J. D, 7, 525


\bibitem[Morton(1991)]{morton91}
  Morton, D. C. 1991,
  ApJS, 77, 119



\bibitem[Penston et al.(1983)]{penston83}
  Penston, M. V., Benvenuti, P., Cassatella, A., et al. 1983,
  MNRAS, 202, 833

\bibitem[Peter \& Judge(1999)]{peter99}
  Peter, H. \& Judge, P. G. 1999,
  ApJ, 522, 1148


\bibitem[Ralchenko et al.(2008)]{ralchenko08}
  Ralchenko, Yu., Kramida, A.E., Reader, J., et al. 2008,
  NIST Atomic Spectra Database (version 3.1.5), National Institute of
  Standards and Technology, Gaithersburg, MD

\bibitem[Robinson(1937)]{robinson37}
  Robinson, H. A. 1937,
  Phys. Rev., 51, 726

\bibitem[Sandlin et al.(1977)]{sandlin77}
  Sandlin, G. D., Brueckner, G. E., \& Tousey, R. 1977,
  ApJ, 214, 898


\bibitem[Schild \& Schmid(1997)]{schild97}
  Schild, H., \& Schmid, H. M. 1997,
  in Physical Processes in Symbiotic Binaries (ed. J. Mikolajewska),
  p.169 

\bibitem[Selvelli \& Bonifacio(2000)]{selvelli00}
  Selvelli, P. L., \& Bonifacio, P. 2000, 
  A\&A, 364, L1

\bibitem[Selvelli et al.(2007)]{selvelli07}
  Selvelli, P., Danziger, J., \& Bonifacio, P. 2007,
  A\&A, 464, 715

\bibitem[Skopal(2007)]{skopal07}
  Skopal, A. 2007,
  New Ast., 12, 597

\bibitem[Smitt et al.(1976)]{smitt76}
  Smitt, R., Svensson, L. \AA, \& Outred, M. 1976, 
  Phys. Scripta, 13, 293


\bibitem[Spyromilio(1995)]{spyro95}
  Spyromilio, J. 1995,
  MNRAS, 277, L59

\bibitem[Thackeray(1950)]{thackeray50}
  Thackeray, A. D. 1950,
  MNRAS, 110, 45

\bibitem[Thackeray(1974)]{thackeray74}
  Thackeray, A. D. 1974,
  MNRAS, 167, 87

\bibitem[Thackeray(1977)]{thackeray77}
  Thackeray, A. D. 1977,
  Mem. RAS, 83, 1

\bibitem[Young et al.(2005a)]{young05a}
  Young, P. R., Berrington, K. A., \& Lobel, A. 2005a,
  A\&A, 432, 665

\bibitem[Young et al.(2005b)]{young05b}
  Young, P. R., Dupree, A. K., Espey, B. R., Kenyon, S. J., \& Ake, T.
  B. 2005b, 
  ApJ, 618, 891

\bibitem[Young et al.(2006)]{young06}
  Young, P. R., Dupree, A. K., Espey, B. R. \& Kenyon, S. J. 2006, 
  ApJ, 650, 1091

\bibitem[Zetterberg \& Magnusson(1977)]{zetterberg77}
  Zetterberg, P. O., \& Magnusson, C. E. 1977,
  Phys. Scripta, 15, 189

\bibitem[Zuccolo et al.(1997)]{zuccolo97}
  Zuccolo, R., Selvelli, P., \& Hack, M. 1997,
  A\&ASS, 124, 425

\end{thebibliography}
\end{document}